\documentclass[twocolumn]{aa}
\usepackage[utf8]{inputenc}
\usepackage{natbib}
\usepackage{graphicx}
\usepackage{placeins}
\usepackage{color}
\usepackage{amsmath}
\usepackage{natbib}
\usepackage{arydshln}
\usepackage{tikz}
\usepackage{endnotes}
\usepackage{hyperref}

\newcommand{\FIG}[1]{#1}

\usepackage{geometry}
\title{Two-fluid implementation in {\tt MPI-AMRVAC}, with applications in the solar chromosphere. }
\author{B. Popescu Braileanu
          \inst{1}
          \and
          R. Keppens \inst{1}}
\titlerunning{Two-fluid equations implementation}
\authorrunning{Popescu Braileanu and Keppens}
\institute{Centre for mathematical Plasma Astrophysics, KU Leuven, 3001 Leuven, Belgium,\\ \email{beatriceannemone.popescubraileanu@kuleuven.be} }
\date{}

\date{Received XXXX; Accepted XXXX}
 
\abstract{
{\it Context.} 
The chromosphere is a partially ionized layer of the solar atmosphere, 
the transition between the photosphere where the gas is almost neutral and the fully ionized corona. 
As the collisional coupling between neutral and charged particles decreases in the upper part of the chromosphere, 
the hydrodynamical timescales may become comparable to the collisional timescale, 
and a two-fluid model is needed.
\\{\it Aims.}
In this paper we describe the implementation and validation of a two-fluid model which simultaneously evolves charges and neutrals, coupled by collisions.
\\{\it Methods.}
The two-fluid equations are implemented  in the fully open-source {\tt MPI-AMRVAC} code. 
In the photosphere and the lower part of the solar atmosphere, where collisions between charged and neutral particles are very frequent, an explicit time-marching would be too restrictive, since for stability the timestep needs to be proportional to the inverse of the collision frequency. This is overcome by evaluating the collisional terms implicitly using an explicit-implicit (IMEX) scheme.
{Out of the various IMEX variants implemented, we focus here on the IMEX-ARS3 scheme, used for all simulations presented in this paper.}
The modular structure of the code allows to directly apply all other code functionality - in particular its automated grid adaptivity - to the two-fluid model.
\\{\it Results.} Our implementation recovers and significantly extends available (analytic or numerical) test results for two-fluid charge-neutral evolutions. We demonstrate wave damping, propagation and interactions in stratified settings, Riemann problems for coupled plasma-neutral mixtures, generalize a shock-dominated evolution from single to two-fluid regimes, and make contact with recent findings on typical plasma-neutral instabilities.
\\{\it Conclusions.}
The cases presented cover very different collisional regimes and our results are fully consistent with related literature findings.
If collisional time and length scales  are smaller than the hydrodynamical scales usually considered in the solar chromosphere,
density structures seen in the neutral and charged fluids are similar, with
the effect of elastic collisions between charges and neutrals being similar to diffusivity.
Otherwise, density structures are different and the decoupling in velocity between the two species increases, and neutrals may e.g. show Kelvin-Helmholtz roll-up while charges do not. The use of IMEX schemes efficiently
avoids the small timestep constraints of fully explicit implementations in strongly collisional regimes.
Adaptive Mesh Refinement (AMR) greatly decreases the computational cost, 
compared to uniform grid runs at the same effective resolution.
}

\begin{document}

\maketitle

\section{Introduction}

The solar chromosphere is a very dynamic layer of the solar atmosphere. It forms the transition between
the dense photosphere with a very low ionization fraction and where the gas pressure is larger than the magnetic pressure, to the very hot, fully ionized corona
dominated by magnetic fields. 
In the corona of the quiet sun, above the transition region located at $\approx$ 2.5 Mm, the plasma is fully ionized \citep{ionFr}.
While a single fluid magnetohydrodynamic (MHD) model for solar plasma applies fully as we reach coronal conditions, the coupling between charged species and the neutrals varies drastically throughout the lower solar atmosphere.
Because the density drops with height, the collisional coupling between neutrals and charges decreases. 
At high collisional frequency, the plasma behaves like a single fluid, so the high collisional frequency near the photosphere again assures an MHD-like behavior. 
On time scales much smaller than  ion-neutral collision time (or equivalently, on length scales much smaller than the mean free path between ions and neutrals), 
the ions and neutrals can be considered  completely decoupled and the two species evolve independently.
However, in an intermediate coupling regime, the
collisions cannot be neglected. 
Partial ionization effects can also be introduced in a single fluid model, through ambipolar diffusion using a generalized Ohm's law.
This approach has been used in many astrophysical contexts for the study of protoplanetary disks \citep{ad-protopl,ad-protopl2} or molecular clouds \citep{ad-mol}.
 A more advanced model for dynamics in partially ionized plasmas must employ 
a fully two-fluid model, where there are separate time evolution equations for neutrals and charged particles (ions, essentially).
This model has also been used in other astrophysical contexts, such as analytical study of
shocks in molecular clouds \citep{1983Draine,Draine1993}.

In the solar atmosphere, the collision frequency between ions and neutrals drops exponentially with height and falls below the ion cyclotron frequency in the chromosphere,
reaching values of $10^2$ s$^{-1}$ at around 2.1 Mm, just below the transition region \citep[see also Figure~1 in ][]{eqkh}.
Above the transition region, in the corona, the plasma is almost fully ionized, however the neutrals still play an important role
in coronal structures with chromospheric origins such as prominences or spicules.
The  study of partial ionization effects is of general importance in astrophysics as the collisions between neutrals 
and charged particles might also have a significant contribution in 
processes related to  accretion disks, interstellar medium, molecular clouds \citep[see also the review of ][]{solerFr}.
Here, we will validate a new implementation of the full two fluid description for plasma-neutral dynamics. A wide variety of previous findings can be used to validate the numerical treatment, going from linear wave dynamics to shocked plasma-neutral evolutions.

Analytical studies of waves using a two-fluid approach showed damping of the waves because of the interaction between charges and neutrals, 
which increases when the collision frequency approaches the wave frequency 
\citep{2011Zaq,2011Zaq2,2013Soler,2013Soler2,2013Zaq,2016Gomez1,2017Gomez1,2018Ballester2,Popescu+etal2018,Popescu2018b}. 
A single-fluid approach, where partial ionization effects are introduced through ambipolar diffusion, 
also showed damping of waves, especially for waves which propagate across the magnetic field \citep{ambi2,beatrice}.
\cite{2011Zaq} shows that these two models to study charge-neutral wave propagation, i.e. the single fluid one with ambipolar diffusion 
and the two-fluid model, give similar results while collisions between ions and neutrals are frequent, however large differences appear
in a weaker coupling regime. {We will present wave applications relevant for the solar chromosphere that relate to these findings.}
Beyond linear wave applications, studies of shocks in a two-fluid approach clearly showed shock substructures that do not appear in the single-fluid assumption \citep{2016Hillier,2019Snow}. 
We will recover these findings, also extended to Riemann problems that already pose numerical resolution challenges in a single fluid MHD approach.

Depending of the hydrodynamical scales considered, collisions between charges and neutrals might decrease or enhance the growth 
of instabilities and modify their evolution.
\cite{Diaz2012} studied analytically the classical Rayleigh-Taylor instability (RTI) in a two fluid approach.
This is a linear study where the authors consider a 1D background consisting of  a uniform magnetic field
and two regions with different uniform densities separated by an interface. A 3D perturbation is applied at the interface.
They showed that the linear growth rate obtained can be one or two orders of magnitude smaller compared to the value in the incompressible single-fluid assumption.
In  simulations of the RTI in a 2.5D geometry, \cite{Popescu3} also find that the interaction between neutrals and charges decreases the linear growth rate of the instability,
the result being similar to \cite{Diaz2012}. 
In the nonlinear phase, the incomplete collisional coupling can be seen visually as a decrease in contrast in the snapshot images, compared to a stronger coupling regime \citep{Popescu3,Popescu4}. 

Indirectly, the presence of neutrals might decrease the growth of RTI because the density contrast at the interface is smaller when the neutrals are considered.
The study of \cite{Arber_2007} of flux emergence in a 3D setup compares the case of fully ionized plasma
to the case of partial ionized plasma, where they include the partial ionized effects through ambipolar diffusion. 
In their simulation, the condition for RTI of having a heavier fluid on top of a lighter one is created 
consistently by the flux emergence. \cite{Arber_2007} show that in a fully ionized plasma,
more chromospheric material is uplifted in the corona, compared to a partial ionized plasma, 
therefore, the RTI is suppressed when the neutrals are included.

On the other hand, in a 2D setup, in different configurations for the magnetic field, with a component parallel to the direction of the perturbation, \cite{DiazKh2013} remark that
neutrals do not feel the stabilizing effect of the magnetic field on the RTI development. 
\cite{2014bKh} used a single-fluid model with ambipolar diffusion and they  found an increase in the growth rate of RTI by 50\% in the simulations that
included the ambipolar term, compared to the pure MHD simulations. 

The inelastic collisions, related to ionization/recombination processes, might also be important.
The neutral fingers resulting from the RTI produced at the interface between a solar prominence thread and corona 
get ionized at the edges while they enter the much hotter corona \citep{Popescu4}.
Even if ionization/recombination did not play any role in the linear growth of the instability \citep{Popescu4}, 
the increase of charged material at the edge of the fingers might impact secondary processes, such as reconnection \citep{Popescu5}. 
For completeness, all relevant
ionization/recombination terms are implemented as discussed below, but none of the tests presented in this paper includes them: most of the reference results from recent literature avoid these effects as well.

In two-fluid simulations of Kelvin-Helmholtz instability (KHI), \cite{ref1} finds that for very short hydrodynamic timescales 
 all the complex motions are occurring in the neutral fluid which is decoupled from the magnetic field and does not feel the Lorentz force.
At larger scales, the fluids become progressively more coupled. 
We will make contact with these KHI simulations, where also the use of Adaptive Mesh Refinement (AMR) and different higher order reconstructions are illustrated.

Besides the many theoretical motivations to develop a two-fluid simulation tool, there are also observational  proofs of the decoupling in velocity between neutrals and charges in the solar atmosphere context.
The theoretical model of \cite{2002Gilbert} showed that neutrals would slip across the magnetic field in solar prominences, and this has been confirmed by observations \citep{2007Gilbert}.
Direct evidence of this decoupling was shown by deducing different velocities from the Doppler shift measured in spectral lines of ions and neutrals observed simultaneously
\citep{2015Khomenko,2016Khomenko,2017Anan,2019Wiehr}.
Further evidence of ion-neutral decoupling  was reported by  \cite{2011delaCruz}, who deduced misalignment in the direction of chromospheric fibrils and the measured magnetic field vector. 
Solar spicules, especially type II, are also highly dynamic structures which evolve on very small timescales, which might reach the same order
of magnitude as the timescale associated with elastic collisions between ions and neutrals. 
\cite{2017Kuzma} studied solar spicules in a two-fluid approach, but did not find significant differences
between the neutral and the charged spicule.
Still, any expected decoupling between neutrals and charges occurs at very small spatial scales and the study of two-fluid effects imposes the need to resolve scales
smaller than the mean free path between ions and neutrals. Therefore,
an extremely high resolution is needed for the study of two-fluid effects, which can effectively be met through the use of AMR \citep{2018Rony}. AMR has been used in many astrophysical contexts to dynamically adjust (increase/decrease) the resolution in certain regions of the grid, and can also be inherited from an external library such as PARAMESH \citep{paramesh}. Here, we implement the two-fluid equations as a new physics module in the {\tt MPI-AMRVAC} code (\url{http://amrvac.org/}), where a native block-structured AMR is implemented in a robust and efficient manner. By design, this module is then directly useful for 1D, 2D and 3D application, in Cartesian or other orthogonal coordinate systems (cylindrical/spherical), although we will here show Cartesian multidimensional setups only.

We present the two-fluid model in Section \ref{sec:model}. Results obtained from numerical simulations which consider 
very different coupling regimes are given in Section \ref{sec:results} and conclusions are listed in Section \ref{sec:concl}.

\section{The two-fluid model}
\label{sec:model}
\subsection{Governing equations}
The two-fluid equations implemented in {\tt MPI-AMRVAC} are \citep[see also][]{Popescu+etal2018}, 
\begin{eqnarray}
\label{eqs:2fl_start}
\frac{\partial \rho_{\rm n}}{\partial t} + \nabla \cdot \left(\rho_{\rm n}\mathbf{v}_{\rm n}\right) = 
S_{\rm n}\,,
 \\
\frac{\partial \rho_{\rm c}}{\partial t} + \nabla \cdot \left(\rho_{\rm c}\mathbf{v}_{\rm c}\right) = 
-S_{\rm n} \,,
\\
\frac{\partial (\rho_{\rm n}\mathbf{v_{\rm n}})}{\partial t} + \nabla \cdot \left(\rho_{\rm n}\mathbf{v_{\rm n}} \mathbf{v_{\rm n}}  + p_{\rm n} \mathbb{I}  \right)
 = \rho_{\rm n}\mathbf{g} +
\mathbf{R}_{\rm n} \,,
\\
\frac{\partial (\rho_{\rm c}\mathbf{v_{\rm c}})}{\partial t} + \nabla \cdot \left[\rho_{\rm c}\mathbf{v}_{\rm c} \mathbf{v}_{\rm c} +  
\left( p_{\rm c} + \frac{1}{2}B^2\right) \mathbb{I} -\mathbf{B}\mathbf{B} \right]\nonumber\\[0.12cm] 
 =\rho_{\rm c}\mathbf{g}  
-\mathbf{R}_{\rm n} \,,
\\
\frac{\partial e^{\rm tot}_{\rm n}}{\partial t}  +  \nabla \cdot \left[ \mathbf{v}_{\rm n} \left(e^{\rm tot}_{\rm n} + p_{\rm n}\right)    \right ]  
 = \rho_{\rm n}\mathbf{v}_{\rm n}\cdot \mathbf{g}  + 
M_{\rm n} \,,
\\[0.15cm]
\frac{\partial e^{\rm tot}_{\rm c}}{\partial t}  +  \nabla \cdot \Big[ \mathbf{v}_{\rm c} \left( 
e^{\rm tot}_{\rm c} +  p_{\rm c} +\frac{1}{2} B^2 \right) - \mathbf{B} (\mathbf{v}_{\rm c} \cdot \mathbf{B}) 
\nonumber\\[0.12cm]
+ \mathbf{v}_{\rm H} B^2  - \mathbf{B} (\mathbf{v}_{\rm H} \cdot \mathbf{B})
 \Big ]   
  = \rho_{\rm c}\mathbf{v}_{\rm c}\cdot \mathbf{g} + \eta J^2  
-M_{\rm n} \,,
\\
\frac{\partial \mathbf{B}}{\partial t} +  \nabla  \cdot \left[ \left(\mathbf{v}_{\rm c} + \mathbf{v}_{\rm H}\right) \mathbf{B}  - 
\mathbf{B} \left(\mathbf{v}_{\rm c} + \mathbf{v}_{\rm H}\right)\right]= \eta \mathbf{J} .
\label{eqs:2fl_end}
\end{eqnarray}
The above set of Eqs.~(\ref{eqs:2fl_start})-(\ref{eqs:2fl_end}) is written for conserved variables: mass densities, momentum and total energies:
\begin{equation}
e^{\rm tot}_{\rm c} = e_{\rm c}+\frac{1}{2}\rho_{\rm c} v_{\rm c}^2 +\frac{1}{2} B^2\,;\quad
e^{\rm tot}_{\rm n} = e_{\rm n}+\frac{1}{2}\rho_{\rm n} v_{\rm n}^2  \,.
\end{equation}

\noindent
Eqs.~(\ref{eqs:2fl_start})-(\ref{eqs:2fl_end}) allow for an external gravitational field with acceleration $\mathbf{g}$, which appears as a source term at the right hand side (RHS) of the equations.

Eqs.~(\ref{eqs:2fl_start})-(\ref{eqs:2fl_end}) are obtained from the multifluid equations, where three species are considered: ions, electrons and neutral particles.
Further, ions and electrons are combined into the charged fluid with density and pressure $\rho_c$ and $p_c$, by assuming that they have the same temperature and that
the center of mass velocity of the charges  is the ions velocity \citep{Popescu+etal2018}. 
The collisional source terms $S_{\rm n}$, $M_{\rm n}$ and $\mathbf{R}_{\rm n}$ at the RHS of 
Eqs.~(\ref{eqs:2fl_start})-(\ref{eqs:2fl_end}) are described next, in subsection \ref{subsec:coll}. 

The multifluid equations, obtained by taking moments of Boltzmann's equation, are closed after the second moment by 
prescribing the pressures
from the ideal equation of state, {hence pressures relate to internal energy densities through}
\begin{equation}
p_i = (\gamma-1)e_i\,,\text{for $i$=n,c}\,.
\end{equation}

The set of equations is closed by further considering the Maxwell's equations.
We assume non-relativistic plasma, where the displacement current is neglected, therefore the current density is defined as  
$\mathbf{J}=\nabla\times \mathbf{B}$. 
The charge neutrality assumption makes Poisson's equation redundant. 
The condition $\nabla\cdot\mathbf{B}=0$ can be handled by many different mechanisms in the {\tt MPI-AMRVAC}  code \citep{tothdivb,multigrid}. 
Non-ideal contributions to the electric field are the Ohmic resistivity, implemented as a source term and the Hall effect, specified by the Hall velocity,
\begin{equation}
\mathbf{v}_{\rm H} = -\frac{\nu_{\rm H}}{\rho_c} \mathbf{J}\,.
\end{equation}
The resistivity and Hall coefficients, $\eta$ and $\nu_{\rm H}$, respectively, are input parameters. {Although implemented, the tests we consider in this paper all relate to ideal two-fluid charge-neutral settings, where the Hall and resistivity contributions are ignored, i.e. $\eta=0=\nu_{\rm H}$.}

The equations in this paper, including the above Eqs.~(\ref{eqs:2fl_start})-(\ref{eqs:2fl_end}), are written in a non-dimensional unit system, where the magnetic permeability $\mu_0$, the
Boltzmann constant $k_B$ and the hydrogen mass $m_H$
are absorbed in the units and do not appear explicitly.
By defining three characteristic quantities (units), for example  number density ($\bar{n}$), temperature ($\bar{T}$) and length ($\bar{x}$),
the units for all other quantities which appear in Eqs.~(\ref{eqs:2fl_start})-(\ref{eqs:2fl_end}) are calculated as,
\begin{equation}
\label{eq:units}
\bar{\rho} = m_H \bar{n}\,,\bar{p} = k_B\bar{n} \bar{T}\,,\bar{v} = \sqrt{\frac{\bar{p}}{\bar{\rho}}}\,,\bar{B}=\sqrt{\mu_0 \bar{p}}\,, \bar{t}=\frac{\bar{x}}{\bar{v}}\,.
\end{equation}
The values {of $\bar{n}$, $\bar{T}$ and $\bar{x}$, as well as of $\mu_0$, $k_B$ and $m_H$ depend on the application considered and the choice of physical unit system,  and both ``SI'' or ``cgs''  choices are available in {\tt MPI-AMRVAC}.}

The temperature is defined by the ideal equation of state,
\begin{equation}
T_i = \mu_i p_i/\rho_i\,, \text{for $i$=n,c}\,.
\end{equation}
We assume a purely Hydrogen plasma, and together with the charge neutrality assumption, this implies that the non-dimensional mean molecular weights are fixed to $\mu_n=1$ and $\mu_c=0.5$. 

Heat conduction (which would be isotropic for neutrals and anisotropic for charges, as implemented in the code) is neglected for all tests presented in this paper, and is hence not listed in the above equations.
The adiabatic constant $\gamma$ is an input parameter and  for the tests presented here is considered to be  $\gamma=5/3$. 

\subsection{Collisional coupling terms}
\label{subsec:coll}
{The RHS source terms $S_{\rm n}$, $\mathbf{R}_{\rm n}$  and $M_{\rm n}$ that balance exactly between neutral and charged species encode both elastic collisions and inelastic contributions (through ionization/recombination).}
The elastic collisions between charges and neutrals include collisions between ions and neutrals and collisions between electrons and neutrals, therefore 
the collisional parameter $\alpha$ is defined so that: $\rho_e \nu_{en} + \rho_i \nu_{in} = \rho_{\rm n} \rho_{\rm c} \alpha$, where
$\nu_{en}$ and $\nu_{in}$ are the collision frequencies between electrons and neutrals and between ions and neutrals, respectively.
The collision frequency between particles $\alpha$ and $\beta$,  as defined by Eq. (\ref{eq:nu_el}) in Appendix \ref{app:collTerms}, 
is  different from the collision frequency  between particles $\beta$ and $\alpha$; $\nu_{\alpha\beta} \ne \nu_{\beta\alpha}$, but $\rho_\alpha \nu_{\alpha\beta} = \rho_\beta\nu_{\beta\alpha}$. 
Because $m_e \ll m_i$, we consider that $\rho_i=\rho_{\rm c}$ and we neglect the collisions between neutrals and electrons. {This makes the collisional parameter simply} $\alpha={\nu_{in}}/{\rho_{\rm n}}$ and its full functional dependence is given by Eq.~(\ref{eq:alpha_el}). In the expression of the collisional parameter $\alpha$ 
the number density cancels out and $\alpha$ depends weakly on the plasma parameters, i.e. on the square root of the temperature average between neutrals and charges.
For this reason and for easier comparison with analytical or other results present in the literature, for the cases presented in this paper we will assume $\alpha$ uniform and constant
and its value will vary throughout the simulations.  

The inelastic collisions are related to ionization and recombination processes and the functional dependence of the ionization/recombination rates $\Gamma^{\rm ion}$, $\Gamma^{\rm rec}$
is given in Appendix by Eqs.~(\ref{eq:gamma_ion}) and (\ref{eq:gamma_rec}), respectively. {None of the tests discussed make use of these inelastic collision terms, but they are implemented and reported here for completeness and for later reference.}

\noindent 
Using the above definitions, the collisional terms in the continuity, momentum and energy equations are: 
\begin{eqnarray}
S_{\rm n} = \rho_{\rm c} \Gamma^{\rm rec} - \rho_{\rm n}\Gamma^{\rm ion}\,,\nonumber\\
\mathbf{R}_{\rm n} = \rho_{\rm c} \mathbf{v}_{\rm c} \Gamma^{\rm rec}  - \rho_{\rm n} \mathbf{v}_{\rm n} \Gamma^{\rm ion} + 
\rho_{\rm n} \rho_{\rm c} \alpha (\mathbf{v}_{\rm c} - \mathbf{v}_{\rm n})\,,\nonumber\\
M_{\rm n} = \frac{1}{2} \rho_{\rm c} v_{\rm c}^2 \Gamma^{\rm rec}  
- \frac{1}{2}\rho_{\rm n} v_{\rm n}^2 \Gamma^{\rm ion} 
+  \frac{1}{2} ({v_{\rm c}}^2 - {v_{\rm n}}^2) \rho_{\rm n} \rho_{\rm c} \alpha
\nonumber\\
+ \frac{1}{\gamma-1} 
\left ( \rho_{\rm c} T_{\rm c} \Gamma^{\rm rec} - \rho_{\rm n} T_{\rm n} \Gamma^{\rm ion} \right) 
\nonumber\\
+\frac{1}{\gamma-1} (T_{\rm c} - T_{\rm n})\rho_{\rm n} \rho_{\rm c} \alpha\,.
\end{eqnarray}
{Note in particular how the momentum exchange via elastic collisions scales with the local velocity difference, and how the related energy exchange scales with squared velocity differences and the temperature difference.}
%

\subsection{Time integration strategy}
%
The collisional terms $S_{\rm n}$, $\mathbf{R}_{\rm n}$ and $M_{\rm n}$ in the continuity, momentum and energy
equations might be stiff when the collisions are very frequent.
For example, the momentum equations, considering only the update due to elastic collisions can be written as
\begin{eqnarray}
\frac{\partial (\rho_{\rm n}\mathbf{v_{\rm n}})}{\partial t} = \nu_{in} (\rho_{\rm c}\mathbf{v_{\rm c}}) - \nu_{ni} (\rho_{\rm n}\mathbf{v_{\rm n}}) \,,\nonumber
\\
\frac{\partial (\rho_{\rm c}\mathbf{v_{\rm c}})}{\partial t} = \nu_{ni} (\rho_{\rm n}\mathbf{v_{\rm n}}) - \nu_{in}  (\rho_{\rm c}\mathbf{v_{\rm c}}) \,,
\label{eq:col_mom}
\end{eqnarray}
where $\nu_{in}$ and $\nu_{ni}$
are the collision frequencies between ions and neutrals and 
between neutrals and ions, respectively, which equivalently write as
\begin{equation}
\label{eq:freqcoll}
\nu_{in} = \alpha \rho_n\,,\quad
\nu_{ni} = \alpha \rho_c\,.
\end{equation}
An explicit implementation of the above terms at the RHS of Eqs. (\ref{eq:col_mom})  is stable if the timestep 
\begin{equation}
\label{eq:expl_restr}
\Delta t \le \frac{1}{\text{max}(\nu_{in},\nu_{ni})}\,. 
\end{equation}
In the solar atmosphere, the very large densities in the lower photosphere and the bottom part of the chromosphere
would impose very small timesteps in an explicit implementation. 
In order to avoid this limitation, the collisional terms 
are evaluated implicitly, so the overall time-stepping employs an implicit-explicit or IMEX scheme, similarly to \cite{Popescu+etal2018}.
The modular structure of the {\tt MPI-AMRVAC} code permits the use of its various, already implemented third, second and first order IMEX schemes. 
In this paper, we will use the IMEX-ARS3 scheme from \cite{ARS3}.

{Within each implicit substep of an IMEX variant, we handle the source evaluations in pairs, and the discussion below has pairs $\mathbf{U}=\left(\rho_n, \rho_c\right)$ or $\mathbf{U}=\left(\rho_n\mathbf{v}_{\rm n}, \rho_c\mathbf{v}_{\rm c}\right)$ and $\mathbf{U}=\left(e^{\rm tot}_{\rm n},e^{\rm tot}_{\rm c}\right)$.}
The implicit update of the collisional terms 
is then based on the fact that the source term for $\partial \mathbf{U}/\partial t=\mathbf{P}(\mathbf{U})$ is proportional to the variables exactly, i.e. $\mathbf{P} = \hat{\mathbf{J}} \cdot \mathbf{U}$.
Because of the weak dependence of the Jacobian matrix $\hat{\mathbf{J}}$ on the variables which evolve in time,
$\hat{\mathbf{J}}$ can be assumed constant during a timestep.
This linearization in time has been considered in many previous implementations by \cite{2008Sakai,2012Toth,2016Hillier,Popescu+etal2018}.

In a multistep semi-implicit scheme the implicit update {by a partial stepsize $\beta\Delta t$} can be written as:
\begin{equation}
\label{eq:impli}
\mathbf{U}^{k+1} = \mathbf{T} + \beta \Delta t \mathbf{P}(\mathbf{U}^{k+1})\,,
\end{equation}
where $\mathbf{T}$ represents the pair of variables after their explicit update. If
$\hat{\mathbf{J}}$ is assumed constant during the step, taking into account that $J_{21}=-J_{11}\,,\quad J_{22}=-J_{12}$, Eq.~(\ref{eq:impli})
 becomes an explicit update:
\begin{eqnarray}
U_1^{k+1} = {T_1} +  \Delta U\,,\nonumber\\
U_2^{k+1} = {T_2} -  \Delta U\,,
\end{eqnarray}
where
\begin{equation}
\Delta U= \beta \Delta t \frac{J_{11} T_1 + J_{12} T_2}{1 + \beta \Delta t (J_{12} - J_{11})}\,.
\end{equation}
\noindent
Therefore, the implicit update can be summarized as:
\begin{equation}
\label{eq:summim}
 \mathbf{U}^{k+1} = \mathbf{T} + f(\beta \Delta t, J_{12}-J_{11})\mathbf{P}(\mathbf{T})\,,
\end{equation}
where
\begin{equation}
\label{eq:du1}
 f(\beta \Delta t, J_{12}-J_{11}) = \frac{\beta \Delta t}{1 + \beta \Delta t (J_{12} - J_{11})}\,.
\end{equation}
{In this generic description, the elements of the matrix $\hat{\mathbf{J}}$} are different for each set of variables: density, momentum, kinetic energy and internal energy:

\noindent
{\bf Continuity:}\\
\begin{equation}
J_{11} = -\Gamma_{\rm ion};\quad J_{12} = \Gamma_{\rm rec}\,.
\end{equation}
{\bf Momentum:}\\
\begin{equation}
J_{11} = -\alpha \rho_c  -\Gamma_{\rm ion};\quad J_{12} = \alpha \rho_n + \Gamma_{\rm rec}\,.
\end{equation}
{\bf Kinetic energy: }\\
the same as for the momentum.\\
{\bf Internal energy:}\\
\begin{equation}
J_{11} = {\mu_n} \left(-\alpha \rho_c  -\Gamma_{\rm ion}\right);\quad 
J_{12} = {\mu_c} \left(\alpha \rho_n + \Gamma_{\rm rec}\right)\,.
\end{equation}
Because of different mean molecular weight for neutrals and charged particles, the Jacobian for the kinetic energy is different to the Jacobian for the internal energy and the 
update of the total energy considers the update of kinetic and internal energy separately.

\noindent
Instead of obtaining the analytical solution of the implicit update starting from Eq.~(\ref{eq:impli}),
\cite{2016Hillier} obtains the analytical solution of the following system of two equations:
\begin{equation*}
\frac{\text{d} \mathbf{U} }{\text{d} t} = \mathbf{P(\mathbf{U})}
\end{equation*}
leading to:
\begin{equation}
\Delta U = \frac{J_{11} T_1 +J_{12} T_2}{J_{12} - J_{11}}\left\{ 1 - \text{exp}\left[ -(J_{12} - J_{11}) \beta \Delta t \right] \right\}\,,
\end{equation}
\noindent
which written in the form defined by Eq.~(\ref{eq:summim}) gives
\begin{equation}
\label{eq:du2}
f(\beta \Delta t, J_{12}-J_{11})=
\frac{1 - \text{exp}\left[ -(J_{12} - J_{11}) \beta \Delta t \right] }{J_{12} - J_{11}} \,.
\end{equation}
The function $f$ defined by  Eq.~($\ref{eq:du1}$) is equal to $f$ defined by Eq.~($\ref{eq:du2}$)
for very small values of the timestep $\Delta t$.
We implemented both variants of the implicit update through the generic function $f(\beta \Delta t, J_{12}-J_{11})$, but only the form described by Eq.~\ref{eq:du1} is considered
in the tests presented here.

\section{Results}
\label{sec:results}
In order to test our new implementation of the two-fluid model in {\tt MPI-AMRVAC} we ran a large series of test simulations. These are presented below, where we go from standard error quantifications, to tests demonstrating the advantages offered by an AMR implementation.

\subsection{Linear waves in uniform media}
We first ran simulations of linear acoustic and Alfv\'en waves in a uniform medium, 
similarly to \cite{Popescu+etal2018}. 
We use the splitting strategy as explained in \cite{Nitin} and similar to that used by the {\textsc{Mancha3D/Mancha3D-2F}} code \citep{Felipe2010}, 
where the equations are  solved for (up to nonlinear) perturbations in density, pressure (energy) and magnetic field.
We show here the convergence tests of Alfv\'en waves in a 1.5D domain (the perturbation and gradients are in the $z$-direction only, but the vectors have
two components: $x$ and $z$), where the values corresponding to the uniform background are:
\begin{equation}
    p_{\rm n0} = 20\,,
    p_{\rm c0} = 10\,,
    \rho_{\rm n0} = 20\,,
    \rho_{\rm c0} = 10\,,
    B_{\rm z0} = 0.4\,.
\end{equation}
The domain is between $z=0.5$ and $z=2.1$. 
After linearizing the two-fluid equations \citep[see Eqs. (36) in ][]{Popescu+etal2018}, where only elastic collisions in the momentum equations are considered, 
assuming solutions of the form
\begin{equation}
\{ v_{\rm nx}, v_{\rm cx}, B_{\rm x1}\} = \{V_n,V_c, B_1\} {\rm exp}[{i(\omega t - k z)}],
\end{equation}
the dispersion and polarization relations are
\begin{eqnarray} \label{eq:alf_an1}
- \rho_{\rm c0} \omega^3 + i  \alpha   \rho_{\rm c0}  ( \rho_{\rm c0} +  \rho_{\rm n0}) \omega^2 +
{B_{\rm z0}}^2  k^2 \omega \nonumber \\ -i  \alpha \rho_{\rm c0} {B_{\rm z0}}^2  k^2 = 0 \,,\nonumber \\
B_1  = - \frac{k {B_{\rm z0}}{V_c}}{\omega}\,, 
{V_n} = \frac{\alpha \rho_{\rm n0}  \rho_{\rm c0} V_c }{i \rho_{\rm n0}  \omega + \alpha \rho_{\rm n0}  \rho_{\rm c0} } \,.
\end{eqnarray}
The above Eqs. (\ref{eq:alf_an1}) are Eqs. (38) and (39) from \cite{Popescu+etal2018}, now written in non-dimensional units.
We choose $k$ corresponding to 5 wavelengths in the domain and the amplitude of the velocity of charges being a fraction of $10^{-3}$ of the background Alfv\'en speed,
so that the waves are in a linear regime.
The analytical solution calculated using these values in the above Eqs. (\ref{eq:alf_an1}) is used as initial condition for the simulations.
The boundary conditions are periodic. 
We considered both weak and strong coupling regimes, where we varied from $\alpha=10^{-1}$, compared to $\alpha=10^4$.

We show here the results only for the semi-implicit temporal scheme IMEX-ARS3, 
used in all the other tests presented in the paper. We adopt
various combination of flux schemes (TVDLF, HLL) and limiters (minmod, cada3, woodward, mp5). 
The Butcher table of the IMEX-ARS3 scheme, which shows the explicit updates, as well as the
implicit updates during a full timestep is \citep{ARS3},
\begin{eqnarray}
    \begin{array}{c|c c c||c c c}
      \delta & \delta & &   & 0 & \delta & \\
      1-\delta & \delta-1 & 2(1-\delta) &    & 0 & 1-2\delta & \delta \\
      \hline
       & 0 & 1/2 & 1/2 &  0  & 1/2 & 1/2
    \end{array}
\,\nonumber\\
\text{with }\delta = \frac{3+\sqrt{3}}{6} \,.
\label{tab:ars3}
\end{eqnarray}
In theory, this IMEX-ARS3 scheme is third order accurate in time \citep{ARS3}, meaning that the numerical error introduced in each timestep is proportional to ${\Delta t}^3$,
where ${\Delta t}$ is the timestep used. {This IMEX scheme has in practice three explicit, and two implicit stages.}

\begin{figure}[!h]
\centering
\FIG{\includegraphics[width=8cm]{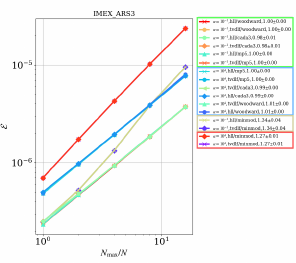}}
\caption{
Error calculated as from Eq. (\ref{eq:err}) as a function of spatial (and temporal) resolution.
$N$ represents the number of points used to discretize the domain.
The points in the plot correspond to $N=$4096 (=$N_{\rm max}$),2048,1024,512,256.
Different combinations of flux schemes (TVDLF, HLL) and limiters (minmod, cada3, woodward, mp5) are shown by different curves, 
some of them overlap and are grouped as indicated in the legend.
}
\label{fig:conv}
\end{figure}

We first run simulations where we vary both temporal ($\Delta t$) and spatial ($\Delta x$) resolution by varying the number of points of the domain
and letting $\Delta t$ to be set by the CFL condition.
For $N \in \{256,512,1024,2048,4096\}$, 
${\Delta t}_N  \in \{3.872 \times 10^{-3}, 1.936\times 10^{-3}, 9.68 \times 10^{-4}, 4.84 \times 10^{-4}, 2.42 \times 10^{-4} \}$, 
Figure \ref{fig:conv} shows the numerical error computed from
\begin{equation}
\mathcal{E}(N) = \max_{i=1}^{N}{\left|u_N\left[ i\right] -  {u_{\rm an}}\left[ i\right]\right| } \,,
\label{eq:err}
\end{equation}
as a function of $N_{\rm max}/N$.  
The variable $u$ considered in the calculation was $v_{\rm cx}$, where
$u_N$ and $u_{\rm an}$ are the numerical 
and analytical solution, respectively, calculated at final time $t_F=1$.
Therefore, the error quantified here is the global truncation error (GTE).
The temporal GTE is the error accumulated after a (large) number of timesteps, which is proportional to $1/\Delta t$, so that for a scheme with order of accuracy $p$,
$\text{GTE}(\Delta t) \propto (\Delta t)^{p-1}$. 
The expression from Eq. (\ref{eq:err}), representing the maximum error in the spatial domain,
also accounts for the accumulated numerical error due to the spatial discretization.
The values of the slopes in Figure \ref{fig:conv} indicate an order of accuracy around 2.
We can see that the normally more diffusive ``minmod'' limiter has a higher convergence rate 
than the other limiters (``cada3'', ``mp5'', ``woodward''), but has a larger error. 
For the same combination of flux schemes and limiters, the error is smaller for smaller values of the coupling parameter $\alpha$ {(i.e. our bright green curve lies below the blue curve, just like the brown curve is always below the red).}
The convergence rate (the slope of the curve) is slightly different for the different values of $\alpha$, and this rate is slightly larger for the smaller value of $\alpha$ for all the combinations, except for those
with a ``woodward'' limiter, where the convergence rate is slightly smaller for the smaller value of $\alpha$.
We are led to conclude that the error we show in Figure~\ref{fig:conv} is dominated by a spatial discretization error {(since a finite volume spatial discretization choice with TVDLF/minmod is at best second order)} and we try to find the 
accuracy of the temporal discretization only.

In order to calculate  the  order of accuracy of the temporal scheme only, we now keep the spatial resolution fixed $N =4096$ and vary the timestep
using a sequence which starts with the value determined by the CFL condition and then, successively, use smaller timesteps by dividing the previous one by a factor of 2, i.e. use the sequence
${\Delta t}_j \in \{2.42 \times 10^{-4},  1.21 \times 10^{-4}, 6.05 \times 10^{-5}, 3.025 \times 10^{-5}, 1.5125 \times 10^{-5} \}$
for $j \in \{0,1,2,3,4\}$.
As we found earlier that the spatial discretization error may dominate,
the calculation of the error using the analytical solution might not converge properly.
For this reason, we now calculate the order of accuracy
by using only the successive numerical solutions, where the spatial discretization error cancels out. We quantify this order as $p$, where:
\endnote{see also Eq.~(3) and Table~1 on website: \href{https://www.csc.kth.se/utbildning/kth/kurser/DN2255/ndiff13/ConvRate.pdf}{https://www.csc.kth.se/utbildning/kth/kurser/DN2255/ndiff13/ConvRate.pdf}} 
\begin{equation}
p=\left\langle \log_2\left( 
\frac{
\sqrt{\sum_{i=1}^{N}{\left(u_j\left[ i\right] -  {u_{j+1}}\left[ i\right]\right)^2 }} 
}{
\sqrt{\sum_{i=1}^{N}{\left(u_{j+1}\left[ i\right] -  {u_{j+2}}\left[ i\right]\right)^2 }} 
}
\right)\right\rangle_{j=0,1,2}\,.
\label{eq:order}
\end{equation}
This also eliminates the differences between the numerical and analytical solution due to nonlinear terms present in the equations solved numerically. Table \ref{tab:ars3} shows values of $p$ for the same combinations of flux schemes/limiters from Figure \ref{fig:conv}.
\begin{table}[!h]
\caption{Quantity $p$ calculated from Eq. (\ref{eq:order}) for the IMEX-ARS3 scheme.}
\begin{center}
\begin{tabular}{c|c|c}
    & $\alpha=10^{-1}$&$\alpha=10^4$\\
    \hline
 TVDLF/minmod   & 2.36 & 2.41 \\
 TVDLF/mp5   & 2.77 & 2.41 \\
 TVDLF/cada3   & 2.74 & 2.41 \\
 TVDLF/woodward   & 2.76 & 2.79 \\
 HLL/minmod   & 2.27 & 2.41 \\
 HLL/mp5   & 2.77 & 2.41 \\
 HLL/cada3   & 2.74 & 2.41 \\
 HLL/woodward   & 2.76 & 2.79 \\
\end{tabular}
\end{center}
\label{tab:conv}
\end{table}
The ``minmod'' limiter now shows  slower convergence, but the dependence on $\alpha$ is the same as in the previous calculation.
The order of accuracy of the temporal scheme is larger than 2, being indeed larger than the order of accuracy of the spatial scheme. 

In summary, the tests 
showed good convergence, of order around 2 for the spatial discretization and close to 3 for the temporal discretization, consistently
with the theoretical estimation.
We observed better overall convergence and smaller error for 
weaker coupling (smaller $\alpha$, where we varied from $\alpha=10^{-1}$, compared to $\alpha=10^4$), 
similarly to \cite{Popescu+etal2018}.
In the strong coupling regime presented here ($\alpha=10^4$), if the collisional terms are implemented explicitly, the timestep is restricted by the collisions,
$\Delta t = 4 \times 10^{-6}$, being three orders of magnitude smaller than the 
timestep used by a semi-implicit scheme for a resolution of 256 points,
making the use of the semi-implicit schemes much more efficient. 
\subsection{Waves through the solar chromosphere}

We then run simulations of waves in a gravitationally stratified atmosphere resembling the solar chromosphere. 
We construct a 1D gravitationally stratified atmosphere for charges and neutrals. The vertical domain is contained between 0.5 Mm and 2.1 Mm. We use the temperature profile 
and the number density of neutrals and charges from the VALC model \citep{valc} and integrate numerically
the  hydrostatic equations. 
The units chosen for the non-dimensionalization are: $\bar{x}=1$~Mm, $\bar{T}=5000 $~K, $\bar{n}=10^{20}$~m$^{-3}$
The value of the collisional parameter is uniform and constant, and was calculated as a spatial average of the coefficient defined by the expression from Eq~(\ref{eq:alpha_el}), 
and in non-dimensional units it has the value $\alpha=1.3\times10^7$. 
\begin{figure}[!htb]
\centering
\FIG{\includegraphics[width=8cm]{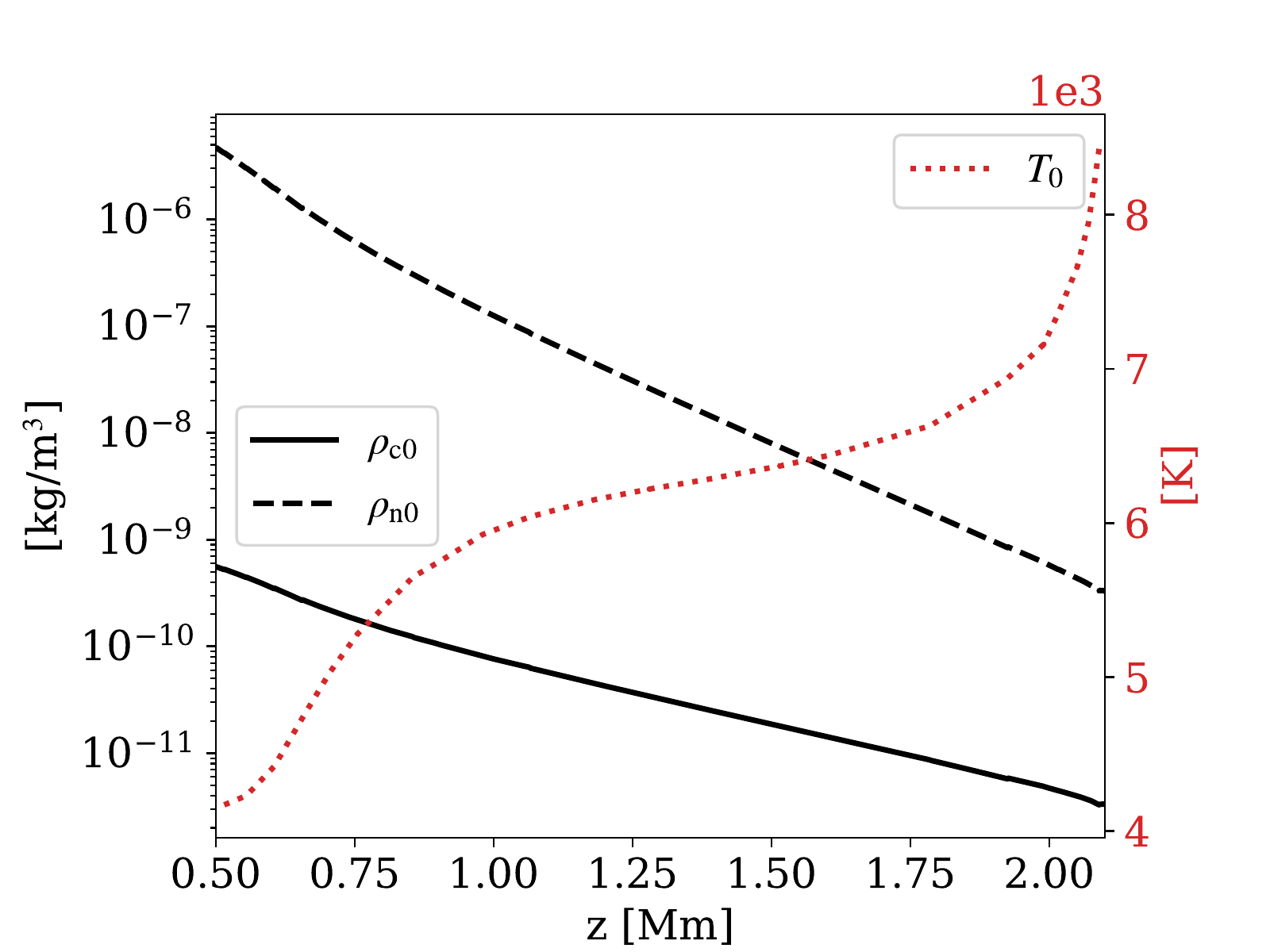}}
\caption{
Equilibrium density of charges (black solid line) and neutrals (black dashed line) on the left axis,
and temperature on the right axis as a function of height.
The equilibrium temperature is the same for neutrals and the charges.
}
\label{fig:equi_strat}
\end{figure}
\begin{figure}[!h]
\centering
\FIG{\includegraphics[width=8cm]{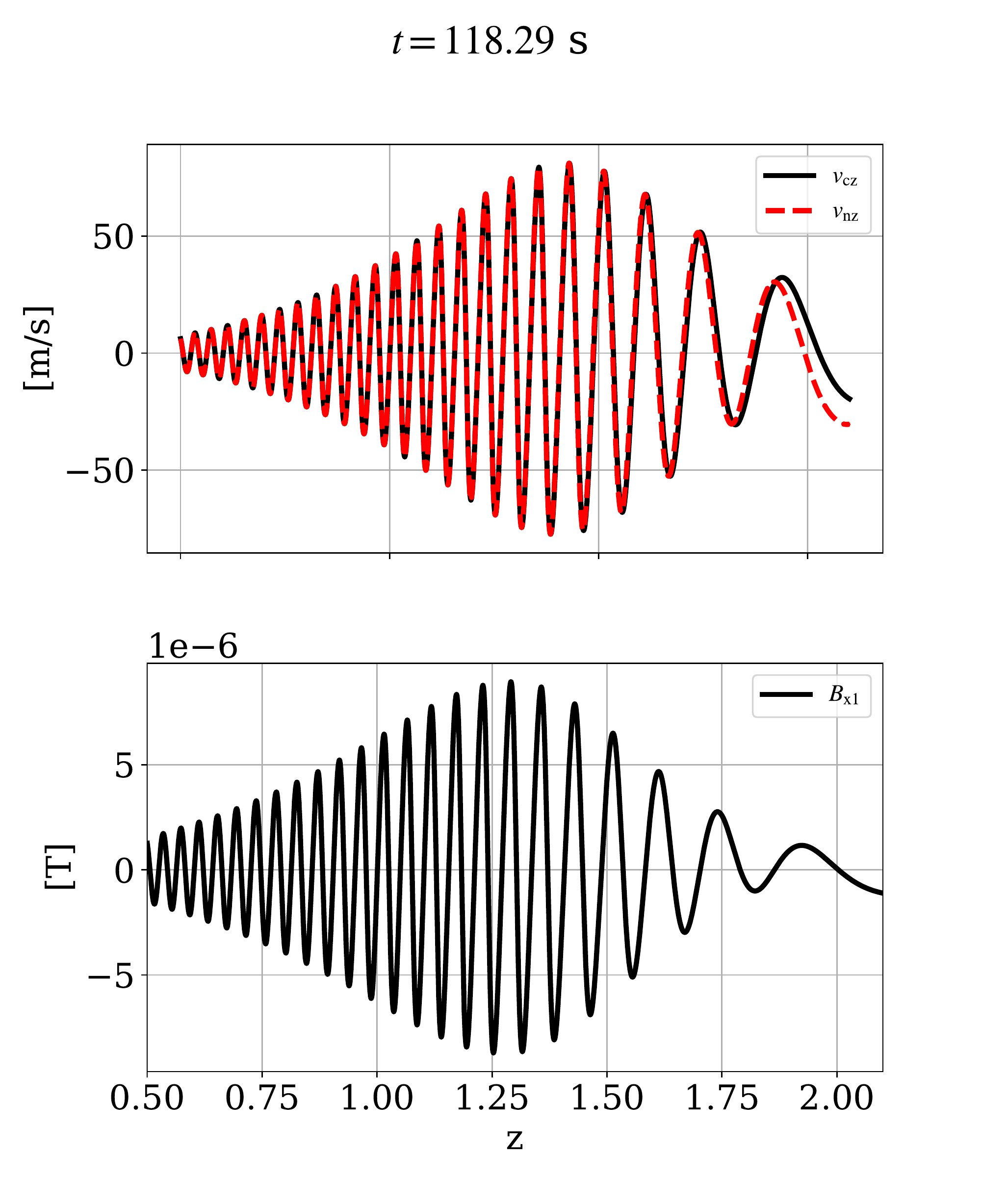}}
\caption{
Snapshot of a 1D fast magneto-acoustic wave in a stratified atmosphere (the horizontal $z$-axis denotes height). Top panel: 
vertical velocity of charges (solid black line) and neutrals (red dashed line).
Bottom panel: perturbation of horizontal magnetic field.
}
\label{fig:maw}
\end{figure}
The densities drop exponentially with height, thus having different coupling regimes at the bottom and the top of the domain. 
Figure~\ref{fig:equi_strat} shows the profiles of densities of neutrals and charges on a logarithmic scale on the left axis 
and the profile of the temperature on the right axis.
Because the scale height is inversely proportional to the mean molecular weight, the scale height of charged particles is 
twice the scale height of the neutral species, therefore the density of the neutrals drops (twice) faster than that of the charges,
this being observed also in the density profiles.
We choose a uniform horizontal magnetic field with magnitude of $\approx 15$~G, so that
the magnetic pressure is equal to the neutral pressure at the middle of the vertical domain, the setup being 
very similar to the setup described by Figure~1 of \cite{Popescu2018b} for their ``S''  magnetic field profile.
We used the same splitting strategy described in \cite{Nitin} where the equilibrium densities, pressure and
magnetic field are assumed to obey a split-off (magneto)hydrostatic configuration whose variations with height are given.
The vertical domain between 0.5~Mm and 2.1~Mm is covered by an uniform grid of 3200 points.

We then use a driver in the bottom ghost cells to launch a fast magneto-acoustic wave with a period of 5 seconds.
The perturbation represents a full linear eigenfunction in all the variables, according to the local analytical solution.
The  amplitude of the velocities is equal to a fraction of $10^{-3}$ of the sound speed at the base of the atmosphere, so that the waves start in the linear regime.
Collisions damp these waves in the upper part of the atmosphere. 
This can be clearly seen in Figure~\ref{fig:maw}, showing velocities and horizontal magnetic field perturbation as a function of height. In the ideal MHD approximation, in a stratified atmosphere where density decreases exponentially, the amplitudes of vertical velocity and of the
perturbation of the original uniform magnetic field should grow exponentially. We observe this amplitude growth in the lower part of the atmosphere where coupling is strong, but they
start to decrease after $z\approx 1.4$ Mm.
We also observe this decoupling as a difference in velocity of charges and neutrals in the upper part,
with slightly larger propagation speed for the charges.
Our result is similar to the result obtained
by~\cite{Popescu2018b}, panel (a) of their Figure~4. 

\subsection{Wave interactions throughout the chromosphere}
\begin{figure*}[!h]
\centering
\FIG{
\includegraphics[width=8cm]{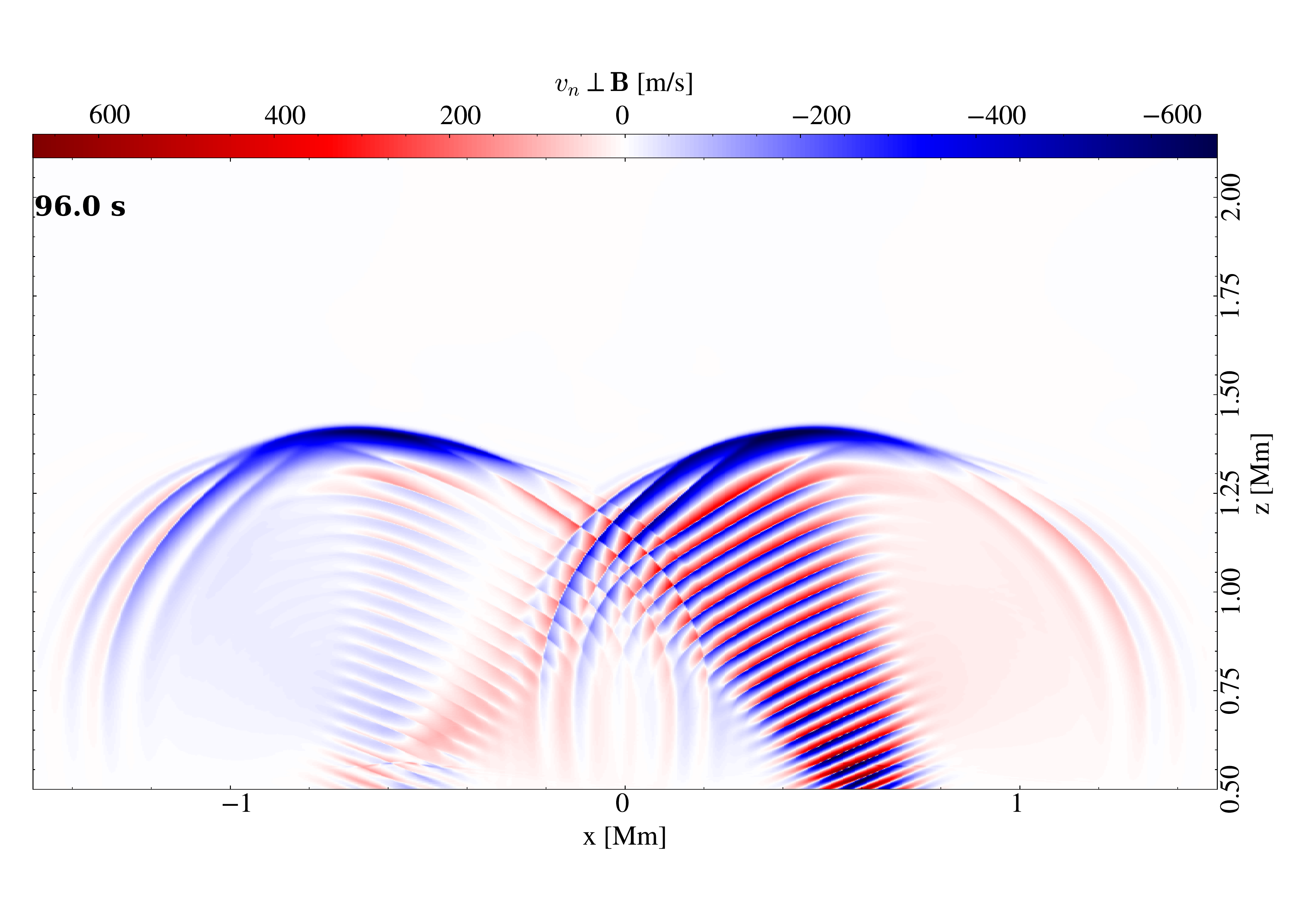}
\includegraphics[width=8cm]{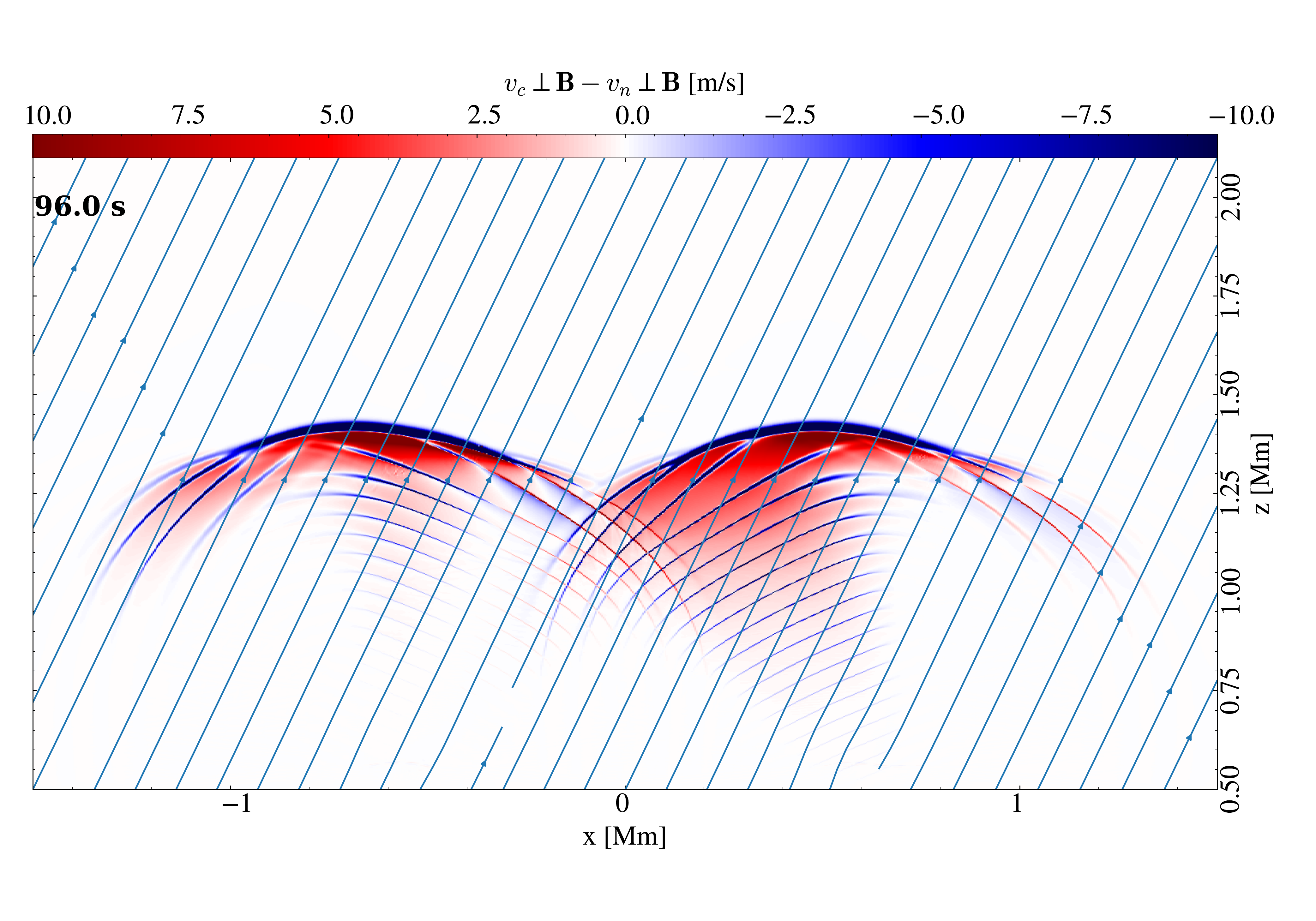}

\includegraphics[width=8cm]{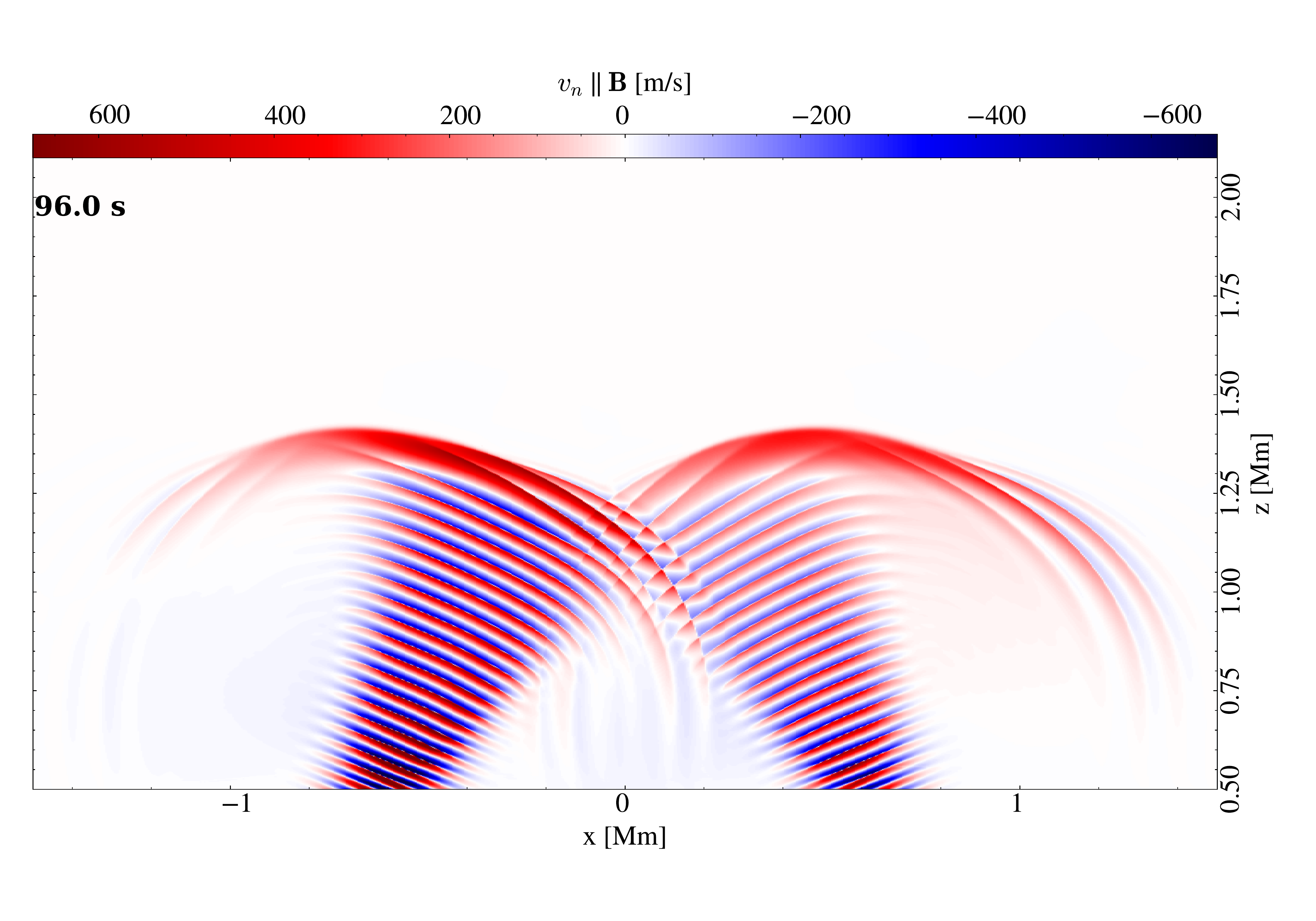}
\includegraphics[width=8cm]{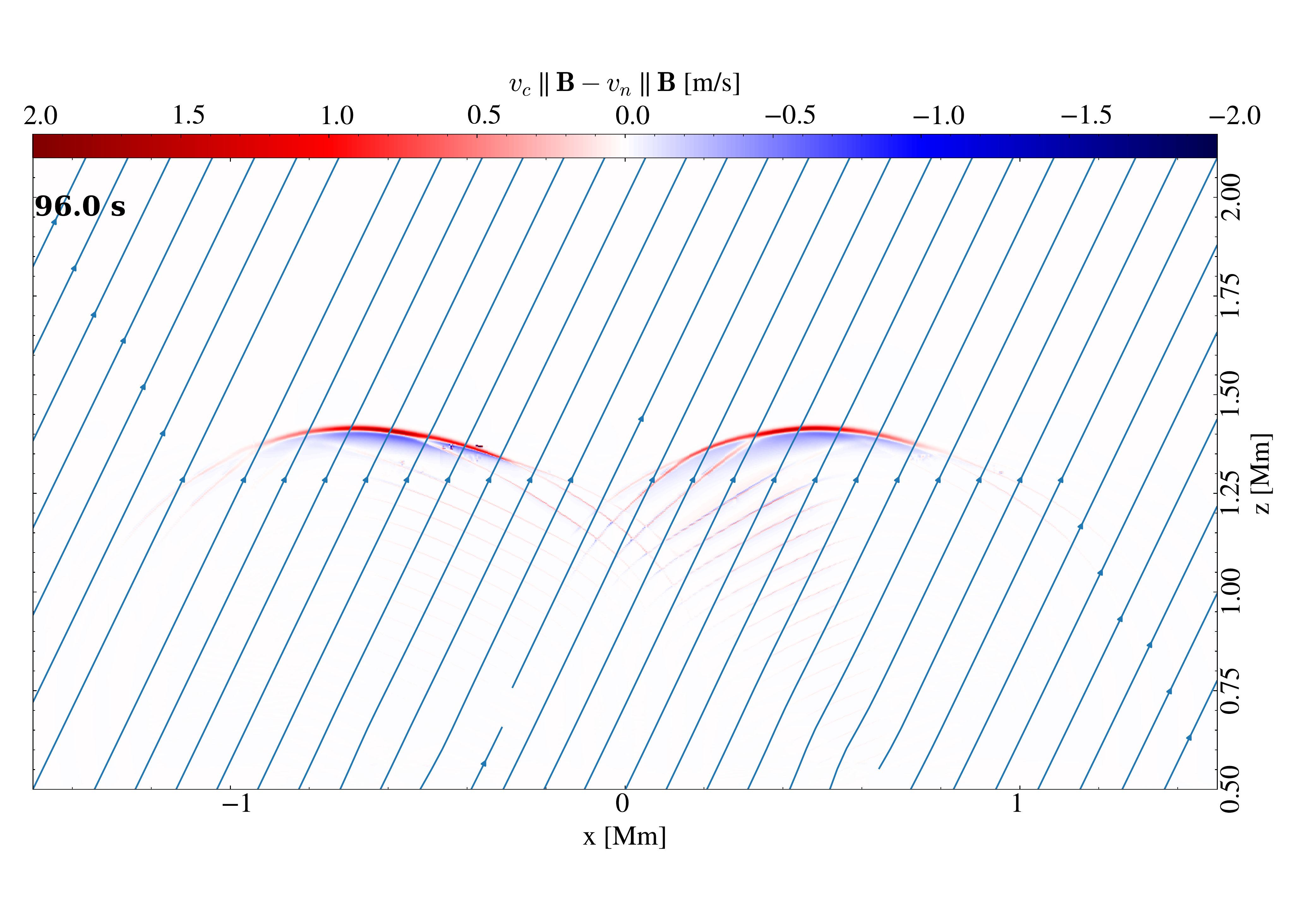}

}
\caption{
2D waves in stratified atmosphere.
Top row: Velocity perpendicular to the magnetic field. Left: neutrals. Right: decoupling (charges - neutrals).
Bottom row: Velocity parallel to the magnetic field. Left: neutrals. Right: decoupling (charges - neutrals).
Note that the limits of the perpendicular (top-right)  decoupling and parallel (bottom-right) decoupling are different.
}
\label{fig:w2f}
\end{figure*} 
\begin{figure*}[!htb]
\centering
\FIG{
\includegraphics[width=8cm,height=6cm]{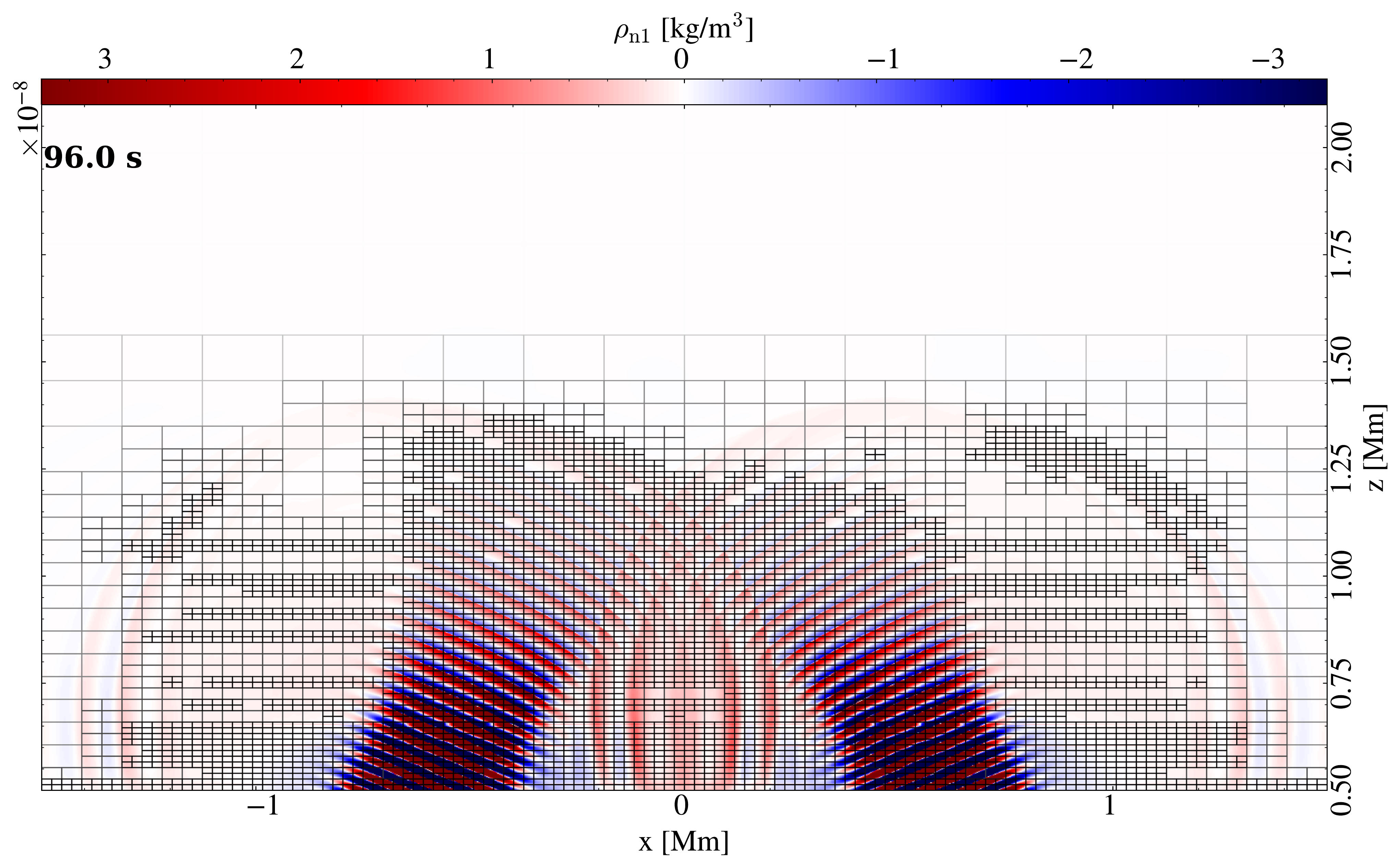}
\includegraphics[width=8cm,height=6cm]{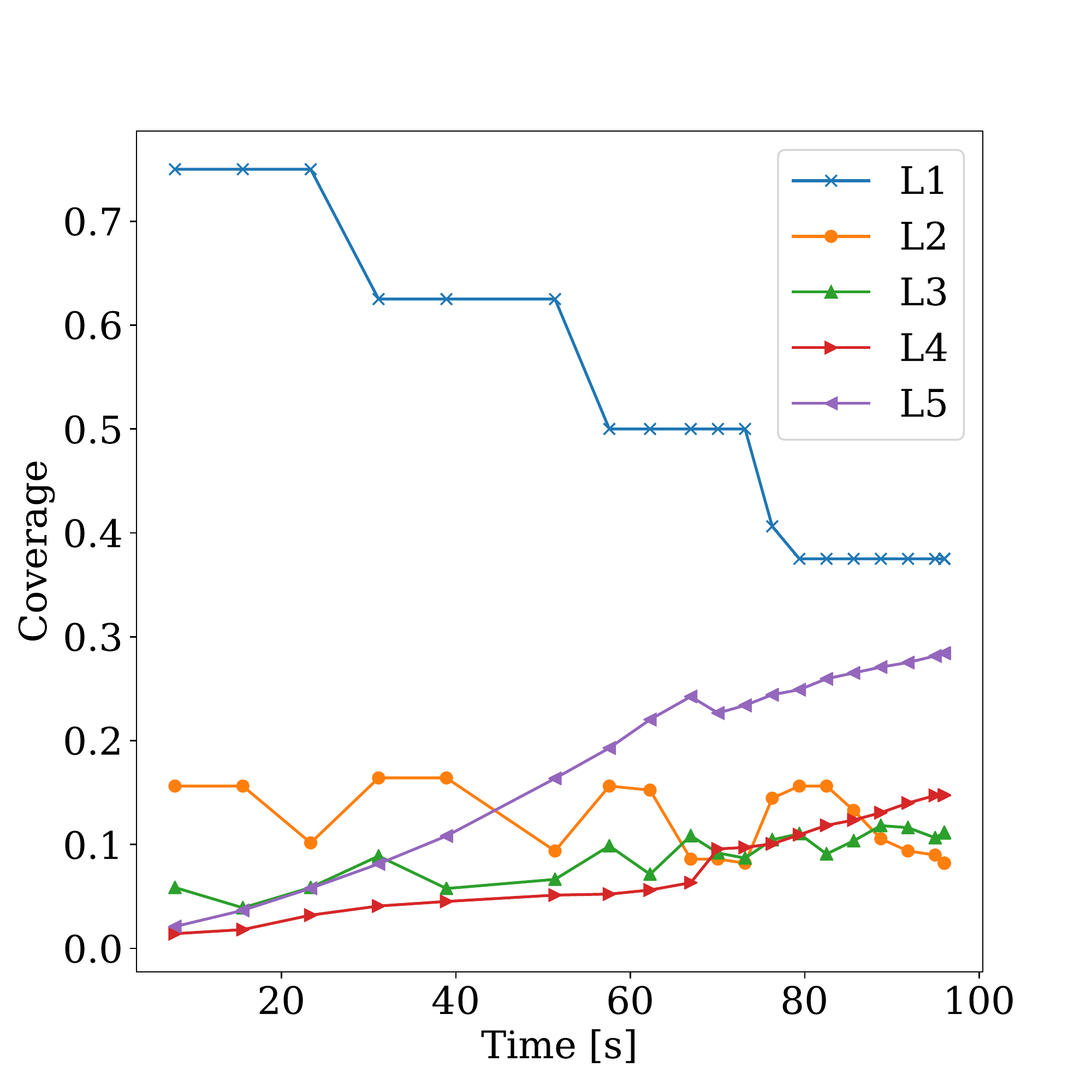}
}
\caption{
Left: Perturbation in the neutral density, where the refinement is shown by overplotting the grids. Right: Coverage of the grid by different levels of AMR as a function of time:
level 1 (blue), level 2 (orange), level 3 (green), level 4 (red), level 5 (violet).
}
\label{fig:w2famr}
\end{figure*}
We then consider a 2D case of the above 1D wave problem, where the magnetic field is still taken to be uniform, but this time inclined with respect to the vertical direction. Its magnitude is still adopted as in the previous 1D case. 
We now launch two fast magneto-acoustic waves with a gaussian shape and different inclination, that will propagate upward and interact throughout the chromosphere.
This is the same experiment that was recently considered by \cite{beatrice} in a single fluid MHD approximation, where partial ionization effects were introduced through ambipolar diffusion.
We use the same period of 5~seconds, but the amplitude of the wave at the base of the atmosphere is now larger by a factor of 100 than in the previous 1D case, 
such that non-linear effects become important.
Our new results, now obtained in the two-fluid approximation, are shown in Figure~\ref{fig:w2f} and can be directly compared to Figure~17 from \cite{beatrice}. 
We show velocity components projected perpendicular (top) or parallel (bottom) to the in-plane inclined field, for the neutrals (left) and in terms of decoupling.
The neutrals can move across the magnetic field lines, contrary to the charged particles. Thus, the decoupling in velocity between charges and neutrals across the field lines 
is much larger than the decoupling along the field lines and this can be seen by comparing our top-right panel to our bottom-right panel of Figure~\ref{fig:w2f}.

Even though the results shown in Figure~\ref{fig:w2f} are at an earlier time that those of Figure~17 from \cite{beatrice}, we can see good agreement for the velocity profiles.
The ambipolar diffusion-based simulation described in \cite{beatrice} is actually more affected by numerical dissipation than the current two-fluid cases shown in Figure \ref{fig:w2f}.
This is because the simulation presented here uses AMR with five levels of refinement  with a base resolution of 1600$\times$400 in $zx$
compared to  the one shown in  \cite{beatrice} where a uniform grid of resolution 3200$\times$800 was used. 
Thus, the effective resolution (i.e. 25600$\times$6400 for the $zx$ domain) is 8 times higher  than that considered in \cite{beatrice}. 
The variable used for the dynamic refinement criterion is the perturbation in neutral density, for which an instant is shown
in the left panel of Figure \ref{fig:w2famr}.  We overplot here the grids and observe that the finer grids are properly located at regions with larger gradient in the neutral density perturbation. 
The grid adjusts and evolves dynamically, and this aspect is shown in the
right panel of Figure \ref{fig:w2famr}, quantifying the grid coverage by the five levels of AMR as a function of time. {This grid coverage is such that the sum over all levels always reaches unity, i.e. our hierarchical grid covers the entire domain.} We can observe that the base level 1 (``L1'') coverage decreases, 
while level 5 (``L5'') increases during the simulation, having a value
close to 0.3 at the end of this simulation.
The fact that the coverage for the highest refinement level (``L5'') remains much lower than unity confirms that
the refinement is done only in specific regions, where needed as specified by our refinement criterion. Therefore the computational cost of this AMR run is obviously much lower than a uniform grid run
with the same cell size as our highest grid level.

\begin{figure*}
\centering
\FIG{\includegraphics[width=16cm]{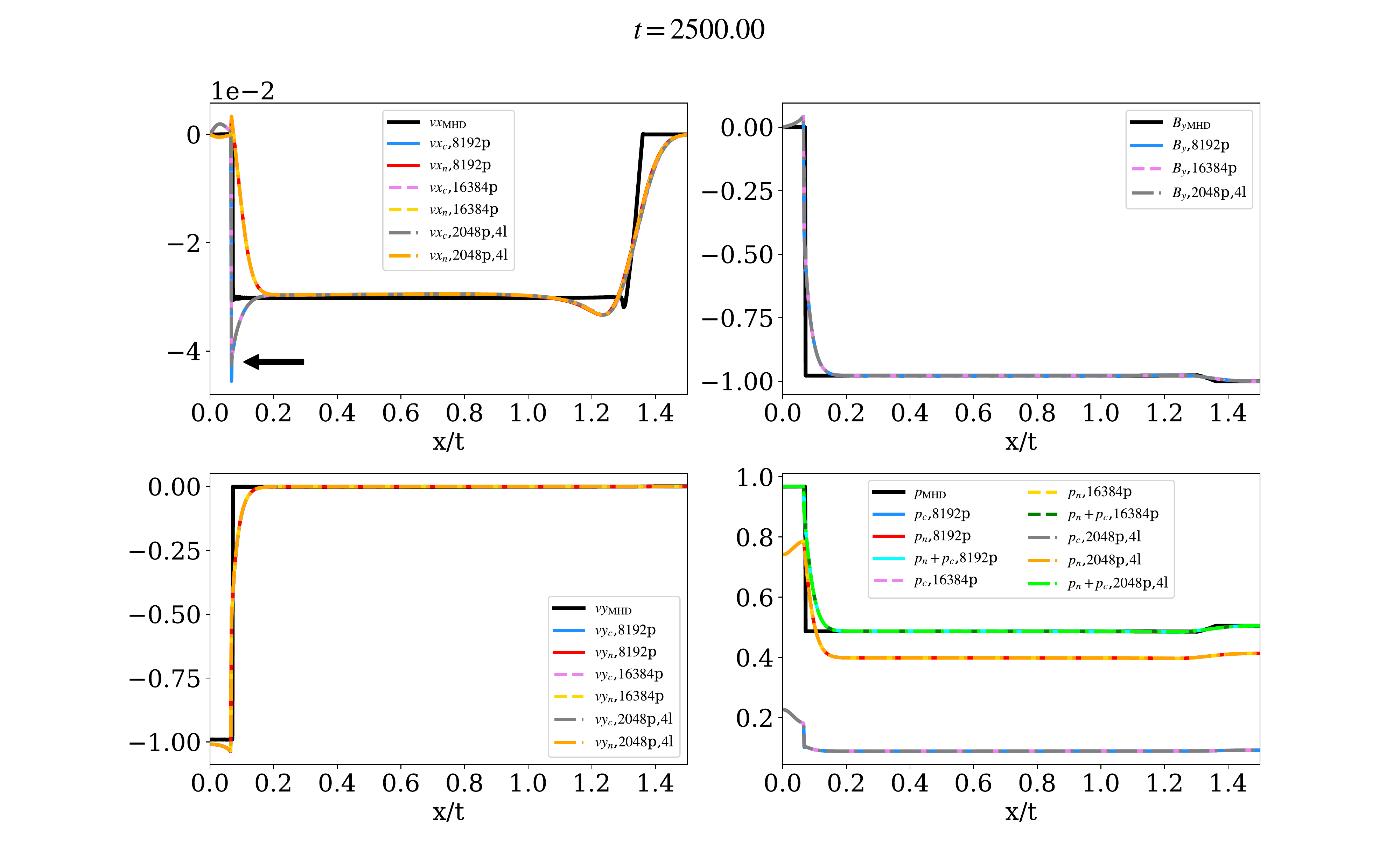}}
\caption{
Self similar solution for the slow MHD and two-fluid shock.
Top left: $x$-velocity; Top right: $B_y$; Bottom left: $y$-velocity; Bottom right: pressure. 
Comparison between runs with different resolutions: uniform grid with 8192 points (solid lines, labeled ``8192p''),
 uniform grid with 16384 points (dashed lines, labeled ``16384''), and AMR with 2048 points base resolution and 4 refinement levels (dot-dashed labeled ``2048p,4l'').
The quantities corresponding to the charges and the magnetic field are plotted with
solid blue (uniform, 8192 points), dashed violet (uniform, 16384 points)  and dash-dotted gray (AMR, 2048 points base resolution, 4 levels);
and for the neutral fluid with 
solid red (uniform, 8192 points), dashed yellow (uniform, 16384 points)  and dash-dotted orange (AMR, 2048 points base resolution, 4 levels).
The reference MHD quantities are shown with solid black lines.
In the bottom left panel the sum of pressures of charged and neutral fluids is also shown as indicated in the legend.
The black arrow in the top left panel indicates the only location where the numerical solution changes visibly for a change in resolution.
}
\label{fig:shocka}
\end{figure*}

\subsection{Two-fluid shock tube}

Here, we repeat the experiment of the slow two-fluid shock in 1.5D geometry done by \cite{2019Snow}, summarized below.
The setup models a shock, which could be  produced by reconnection, where the interface of discontinuity is considered to be at the bottom of the physical domain, $x=0$.
The top of the physical domain is located at $x=4\times 10^3$.
The left and right states of the MHD variables are \citep{2019Snow}:
\begin{eqnarray}
\left( \rho, {v_x}, {v_y}, p, {B_x}, {B_y} \right)_L = \left( \rho_0,0,0,p_0,B_{\rm x0}, -B_{\rm y0} \right)\,,\nonumber\\ 
\left( \rho, {v_x}, {v_y}, p, {B_x}, {B_y} \right)_R = \left( \rho_0,0,0,p_0,B_{\rm x0}, B_{\rm y0} \right)\,,
\end{eqnarray}
where
\begin{eqnarray}
\rho_0=1\,,B_{\rm x0} = 10^{-1}\,,B_{\rm y0} = -1\,,
p_0=\frac{1}{2}(B_{\rm x0}^2+ B_{\rm y0}^2)\,.
\end{eqnarray}
The left state is specified in the bottom ghost cells by the boundary conditions.
Symmetric boundary conditions imply zero gradient and keep
the values of the density, pressure, $y$-component of the velocity and  $x$-component of the magnetic field  the same at the left and the right
of the interface of discontinuity. $B_x$ is uniform and constant throughout.
For the other variables, the $x$-component of the velocity and $y$-component of the magnetic field at the bottom boundary are antisymmetric.
In a 2D setup, this reversal of the magnetic field represents a local current sheet, which is further liable to reconnection. 
The inflow ($x$-component) velocity towards the current sheet is zero. 
In the 1.5D setup done here, we only focus on the outcome of the corresponding Riemann problem.
The top boundary conditions are symmetric for all the variables. 

A MHD uniform setup is generally transformed into a two-fluid setup:
\begin{eqnarray}
\label{eq:mhd-2fl}
    \rho_c = \frac{\xi_i}{\xi_i + 1} \rho\,, &  \rho_n = \frac{1}{\xi_i + 1}\rho\,, \nonumber\\
    p_c = \frac{2 \xi_i}{2  \xi_i + 1} p\,,  &  p_n = \frac{1}{2 \xi_i + 1} p\,,\nonumber\\  
    v_c=v_n=v\,, &
\end{eqnarray}
where $\xi_i$ is the ionization fraction. 
We used the same value $\xi_i=0.1$ from \cite{2019Snow} for this two-fluid setup.

Figure~\ref{fig:shocka} can be directly compared to Figure~4 of \cite{2019Snow}, and shows velocities, $B_y$ variation and pressure variation at time $t=2500$.
Detailed comparison shows a very good agreement. Our figure shows results for a pure MHD run (black solid lines) as they compare to two-fluid runs of different overall resolutions.
In this test we demonstrate the effect of resolution and did the test using uniform grids of 8192 and 16384 points. Visually the solutions for these two resolutions superpose, except for a small undershoot in the charged fluid horizontal velocity for lower resolution, which is
marked in the top-left panel of Figure~\ref{fig:shocka}. We also run simulations using AMR with a base resolution of 2048 points and four levels of refinement,
thus having the same effective resolution as the simulation with 16384 points in a uniform grid.
We observe that AMR and uniform grid results are visually identical, including the undershoot detail shown in our top-left panel.
However, the computational cost is of course smaller when refinement is used. The total computational time for the base resolution of 2048 points and four levels of refinement
is half that of  the case of the uniform grid with 16384 points: 97.09 s compared to 196.587 s, both simulations being run under the same conditions.

\subsection{Shock tube in 1.75D}
\begin{figure*}
\centering
\FIG{\includegraphics[width=16cm]{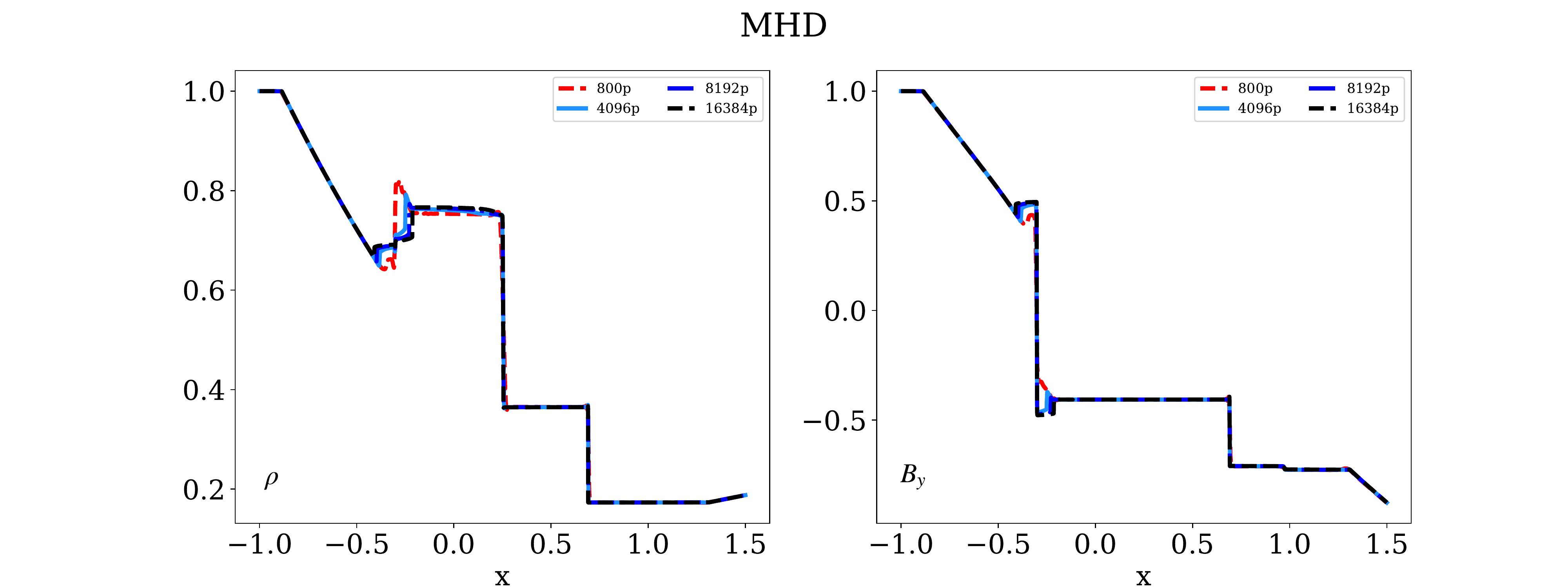}}
\FIG{\includegraphics[width=16cm]{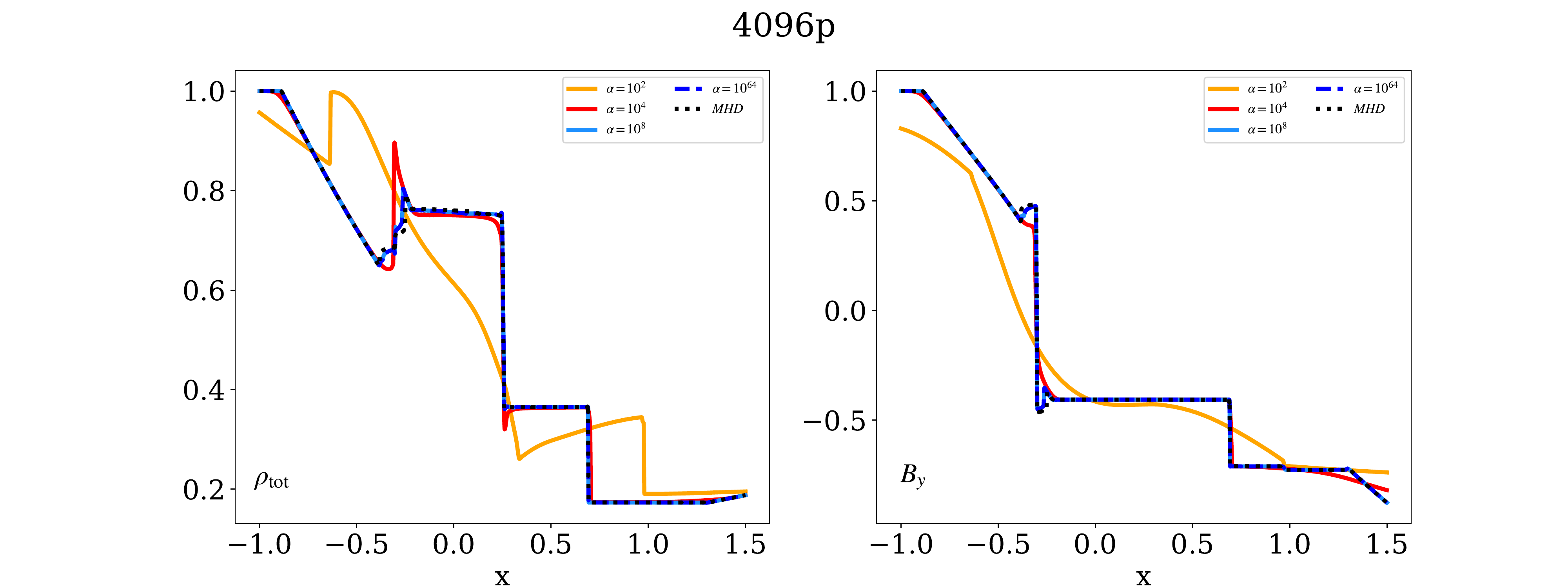}}
\caption{
Solution of the 1.75D shock tube problem at time t=0.5.
Top row: MHD reference result with different resolutions: 800 points (dashed red), 4096 points (solid light blue),
8192 points (solid dark blue), 16384 points (dashed black).
Bottom row: two-fluid model for different values of the collisional parameter: $\alpha=10^2$ (solid orange),
$\alpha=10^4$ (solid red), $\alpha=10^8$ (solid light blue) and $\alpha=10^{64}$ (dashed dark blue).
The reference MHD solution is also shown with a dotted black line.
Left columns: density. For the two-fluid case in the bottom left panel, the sum of densities of neutrals and charges is shown. Right columns: $B_y$.
}
\label{fig:shockc}
\end{figure*}

Collisions affect the scales below the mean free path between neutrals and charges, with their effect being similar to diffusivity for hydrodynamic scales larger than the collisional scale.
{Here, we test a shock tube problem that already poses a challenge to any modern MHD code, as introduced by \cite{torrh} and mentioned in Chapter 20 of \cite{hans2019}} where it is shown that modern shock-capturing discretization
converge to a wrong solution for a resolution lower than several 1000 grid points. 
For larger resolutions the solutions converge to the true and unique solution containing rotational discontinuities (rather than showing a compound wave structure). 
The domain is between $x=-1$ and $x=1.5$ and is covered by an uniform grid with different resolutions, as specified below.
The initial condition in the MHD approximation are:
\begin{eqnarray}
\label{eqs:shockc}
\rho_L = 1\,,\rho_R = 0.2\,,p_L=1\,,p_R=0.2\,,\nonumber\\
{B_y}_L=1\,,{B_y}_R=\text{cos}(3)\,,{B_z}_L=0\,,{B_z}_R=\text{sin}(3)\,,\nonumber\\
{B_x}_L={B_x}_R=1\,,
\end{eqnarray}
where subscripts $_L$ and $_R$ indicate the regions at the left ($x<0$) and right ($x>0$) of the interface located at $x=0$.
The velocities are zero initially.
The boundary conditions are symmetric for all the variables.
The top panels of Figure~\ref{fig:shockc} can be compared to Figure~20.14 from \cite{hans2019}, where we
show pure MHD solutions  at time $t=0.5$, obtained with different resolutions. We can observe that, indeed, the simulation which used
a resolution of 800 points does not capture correctly the region around $x=-0.25$, but a resolution of 4096 points
gives results similar to the expected analytic solution of this Riemann Problem. Further increasing the resolution to
8192 and 16384 points does not show visible changes in the numerical solution.

We then run the same test with the two-fluid model, using Eqs.~(\ref{eq:mhd-2fl}) and $\xi_i=0.1$ to setup all states initially. Here, we keep the same resolution in all the two-fluid simulations, 
namely  4096 points,  which showed the correct numerical solution in the MHD case, but vary the value of the collisional parameter $\alpha$. 
These results are shown in the bottom panels of Figure~\ref{fig:shockc}, along with the reference MHD solution.
We observe that for larger values of $\alpha$ the two-fluid solution converges to the MHD solution. For both largest values of $\alpha$ considered here, 
$\alpha = 10^8$ and $\alpha = 10^{64}$, the two-fluid solution is almost indistinguishable from the MHD solution.  
For the smaller values of $\alpha$, $\alpha=10^2$ and $\alpha=10^4$, 
the two-fluid solution resembles the MHD solution of the run with 800 points, which showed the (erroneous) compound structure.
Therefore in this regime, it seems that the incomplete collisional coupling  has the same effect as numerical diffusivity.
The mean free path between charges and neutrals and between neutrals and charges could be approximated by: 
\begin{equation}
\label{eq:lambda}
\lambda_{in} = \frac{c_f}{\nu_{in}}\,,
\lambda_{ni} = \frac{c_f}{\nu_{ni}}\,,
\text{where } c_f=\sqrt{c_A^2 + c_s^2}\,,
\end{equation}
with the collision frequencies $\nu_{in}$ and $\nu_{ni}$ defined in Eq~(\ref{eq:freqcoll}), and
$c_s$ and $c_A$ being the sound and Alfv\'en speed, respectively, calculated using the total pressure and density.
For the ionization fraction considered here,
the mean free path between neutrals and charges is larger than that between charges and neutrals.
The hydrodynamical scale has the same order of magnitude as the length of the horizontal domain, i.e. $L_x=2.5$.
For the value of $\alpha=10^2$ the mean free path between neutrals and charged particles
is two orders of magnitude smaller than $L_x=2.5$, 
suggesting that neutrals and charges are well coupled. 
In this regime, the incomplete coupling resembles numerical diffusivity.
In weaker coupling regimes, however, the two-fluid solution might be very different from the MHD solution.

\subsection{Shock-interactions in 2D: Orszag-Tang in two-fluid settings}
\begin{figure}[!htb]
\centering
\FIG{
\includegraphics[width=4cm]{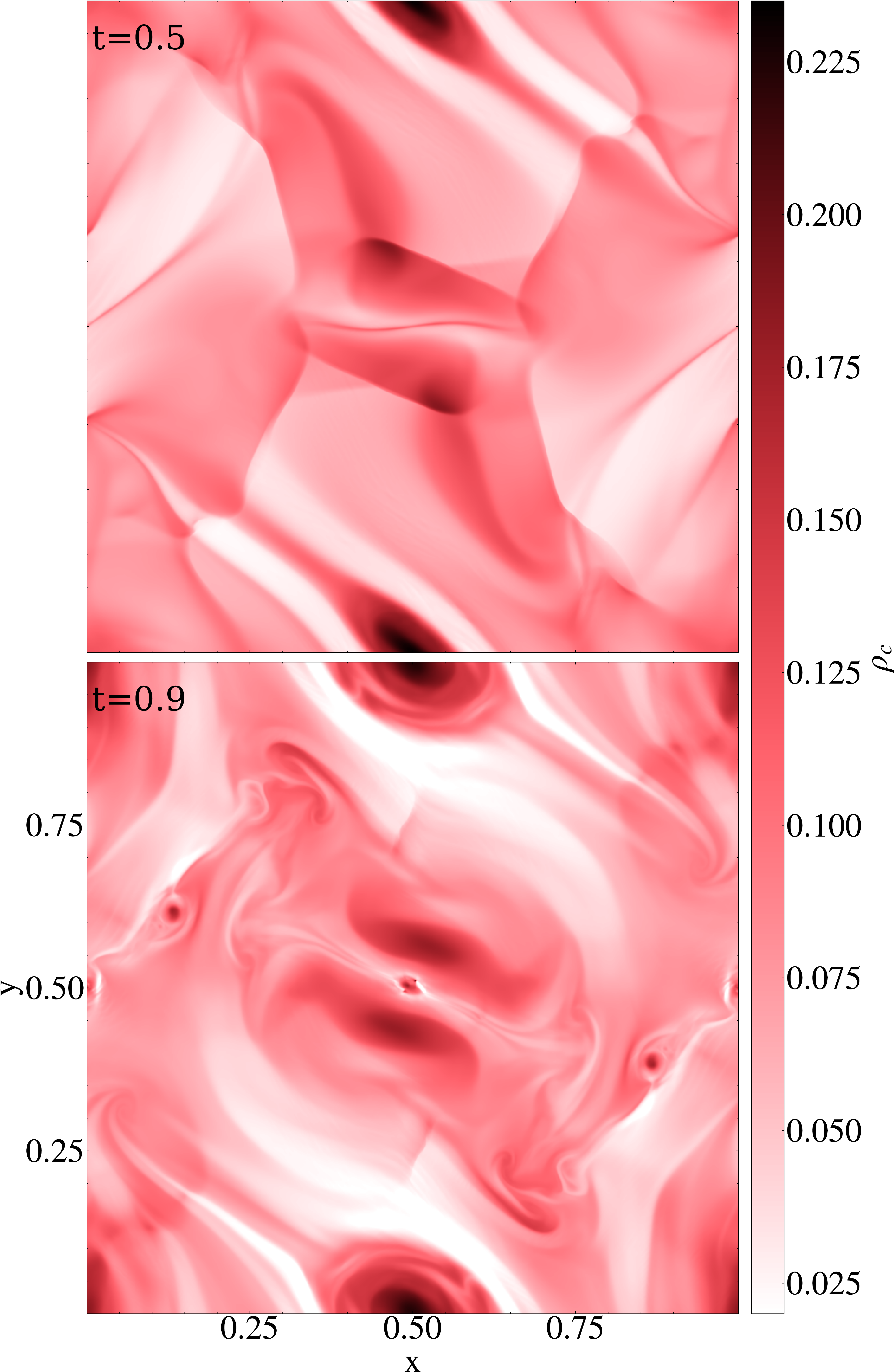}
\includegraphics[width=4cm]{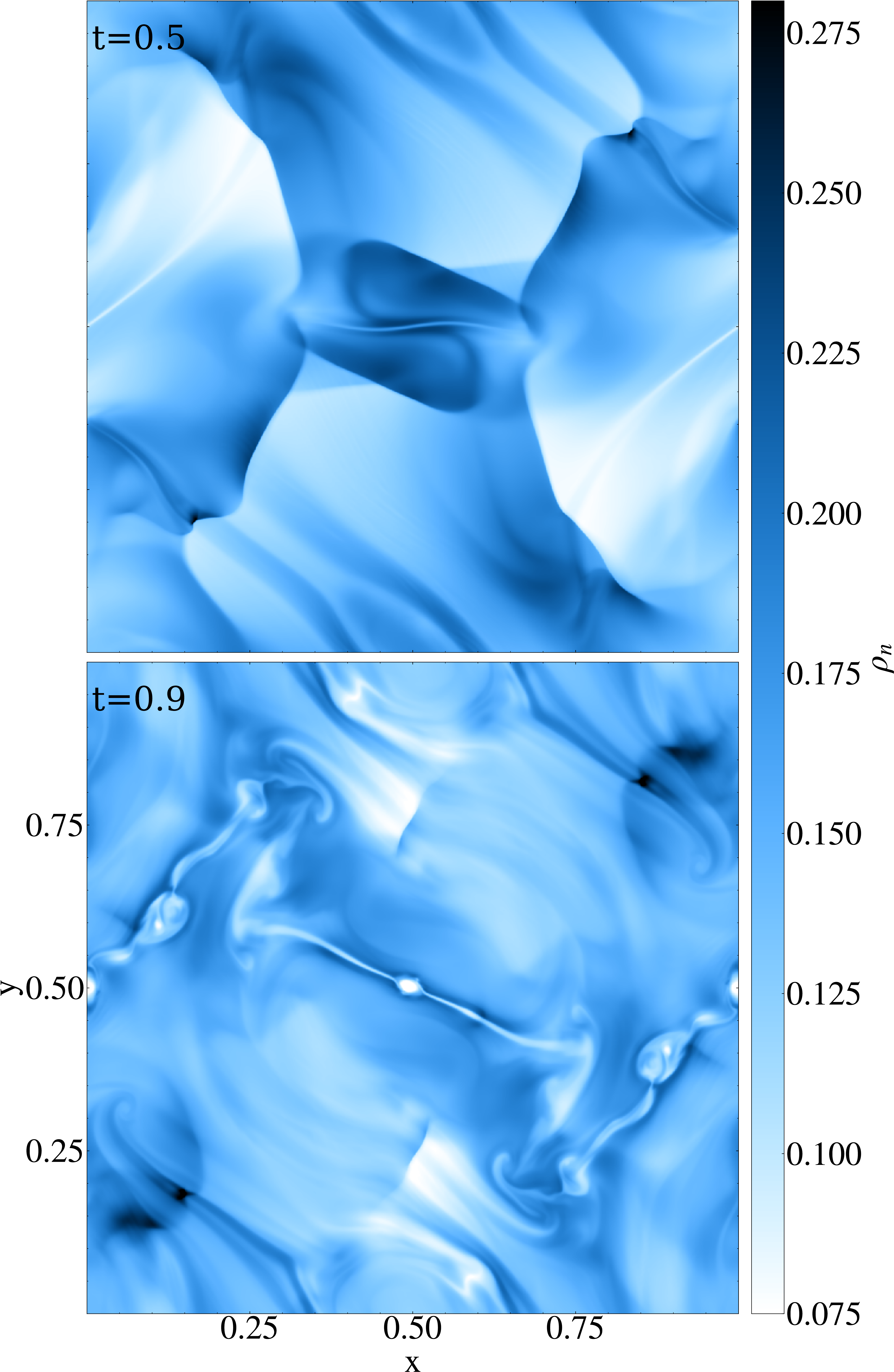}
}
\caption{
Snapshots of densities for the two-fluid model where $\alpha=10^3$.
Left column: charges; Right column: neutrals.
Top row: t=0.5; Bottom row: t=0.9.
}
\label{fig:ors1}
\end{figure}
\begin{figure}[!htb]
\centering
\FIG{
\includegraphics[width=4cm]{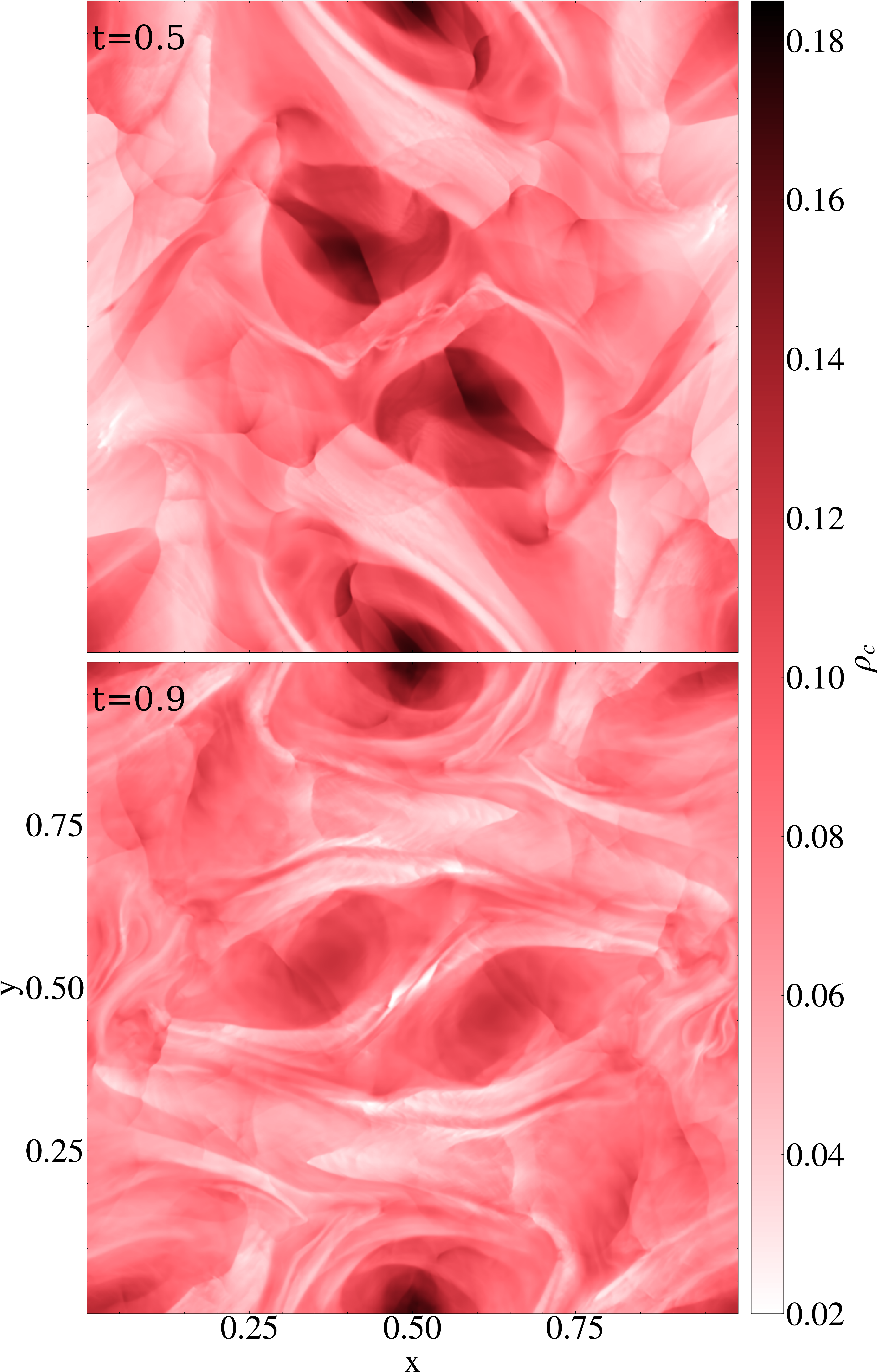}
\includegraphics[width=4cm]{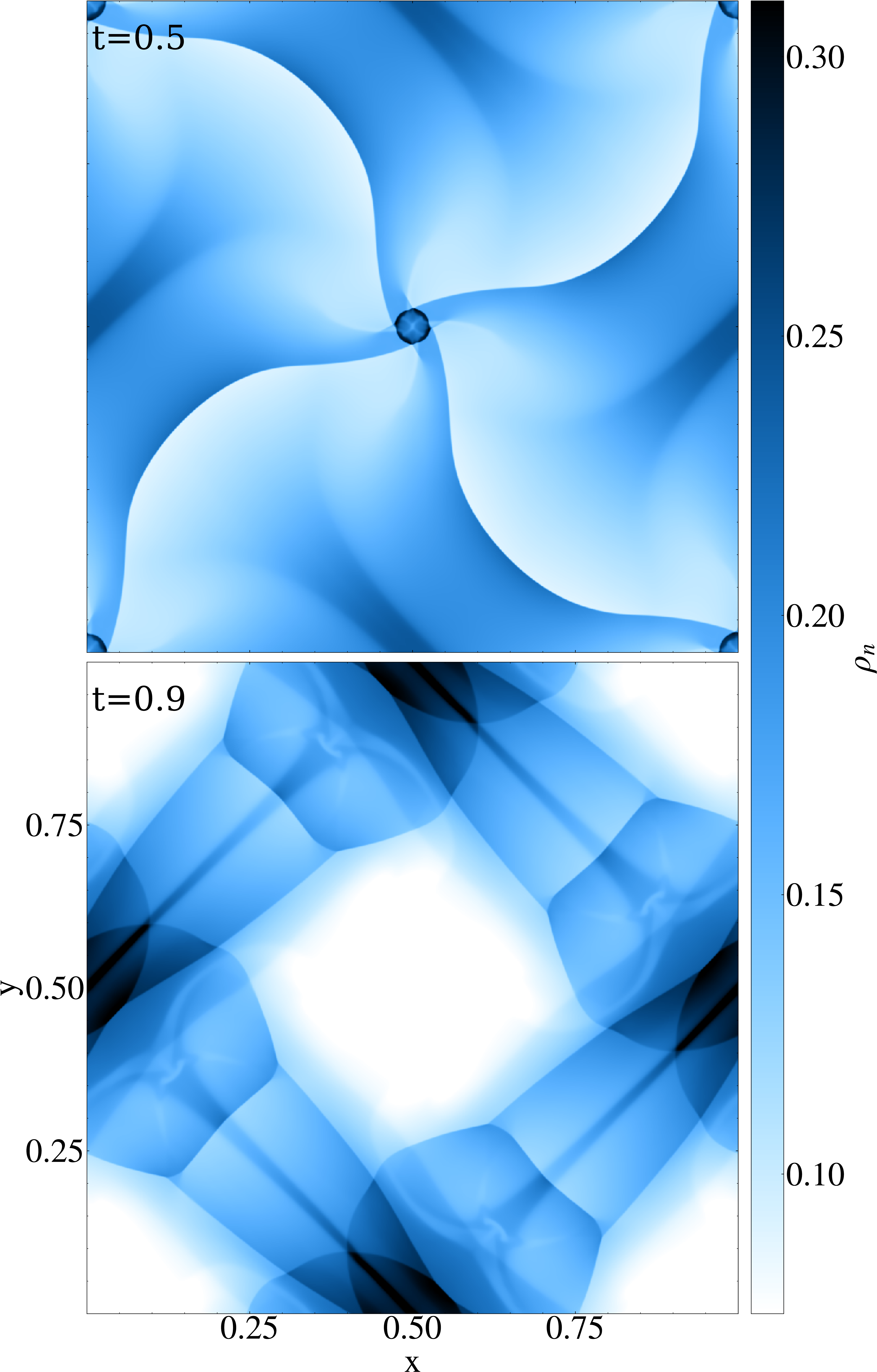}
}
\caption{
Same as Figure \ref{fig:ors1}, for $\alpha=1$.
}
\label{fig:ors2}
\end{figure}

When the coupling regime is weaker, i.e. the collisional scales become similar or larger than the hydrodynamical scales, 
as the neutrals do not feel the magnetic field, 
the two-fluid solution might be very different from the MHD
solution for both neutrals and charges.
In order to see how the code behaves for different regimes of collisional coupling in 
simulations where 2D shock fronts form spontaneously and interact, we do simulations of the compressible version of the
Orszag-Tang test \citep{orszag1,orszag2}. 
The setup for the MHD approximation on a unit square domain is 
\begin{eqnarray}
\rho_0=\frac{25}{36 \pi}\,,\quad p_0=\frac{5}{12 \pi} \,,\nonumber\\
B_{\rm x0} = -B_0 \text{sin}(2 \pi y)\,,\quad B_{\rm y0} = B_0 \text{sin}(4 \pi x)\,,\nonumber\\
\text{where } B_0=\frac{1}{\sqrt{4 \pi}}\,,\nonumber\\
v_{\rm x0} = -\text{sin}(2 \pi y)\,,v_{\rm y0} = \text{sin}(2 \pi x)\,,
\end{eqnarray} 
is extended to a two-fluid setup by using Eqs. (\ref{eq:mhd-2fl}) and the ionization fraction $\xi_i = 0.5$.
We used a base resolution of $512^2$ and 3 levels of refinement. The variable used in the refinement criterion is the magnitude of the decoupling in velocity.
Figure~\ref{fig:ors1} shows snapshots of charged (left) and neutral (right) density for the case of a strongly coupled regime where $\alpha=10^3$, at two different times.
At the ionization fraction chosen, the mean free path between charges and neutrals is the same as the mean free path between 
neutrals and charges, and for the value of $\alpha=10^3$, $\lambda_{in}=\lambda_{ni}=5.28\times10^{-3}$, much smaller than the length of the domain.
We clearly observe similar structures in neutral and charged densities, even up to the very nonlinear times where islands form due to numerical reconnection on the strong and 
localized current sheets (see the snapshots at $t=0.9$ in Fig.~\ref{fig:ors1}). This result is very similar to a pure ideal MHD run. 

\begin{figure}[!htb]
\centering
\FIG{
\includegraphics[width=8cm]{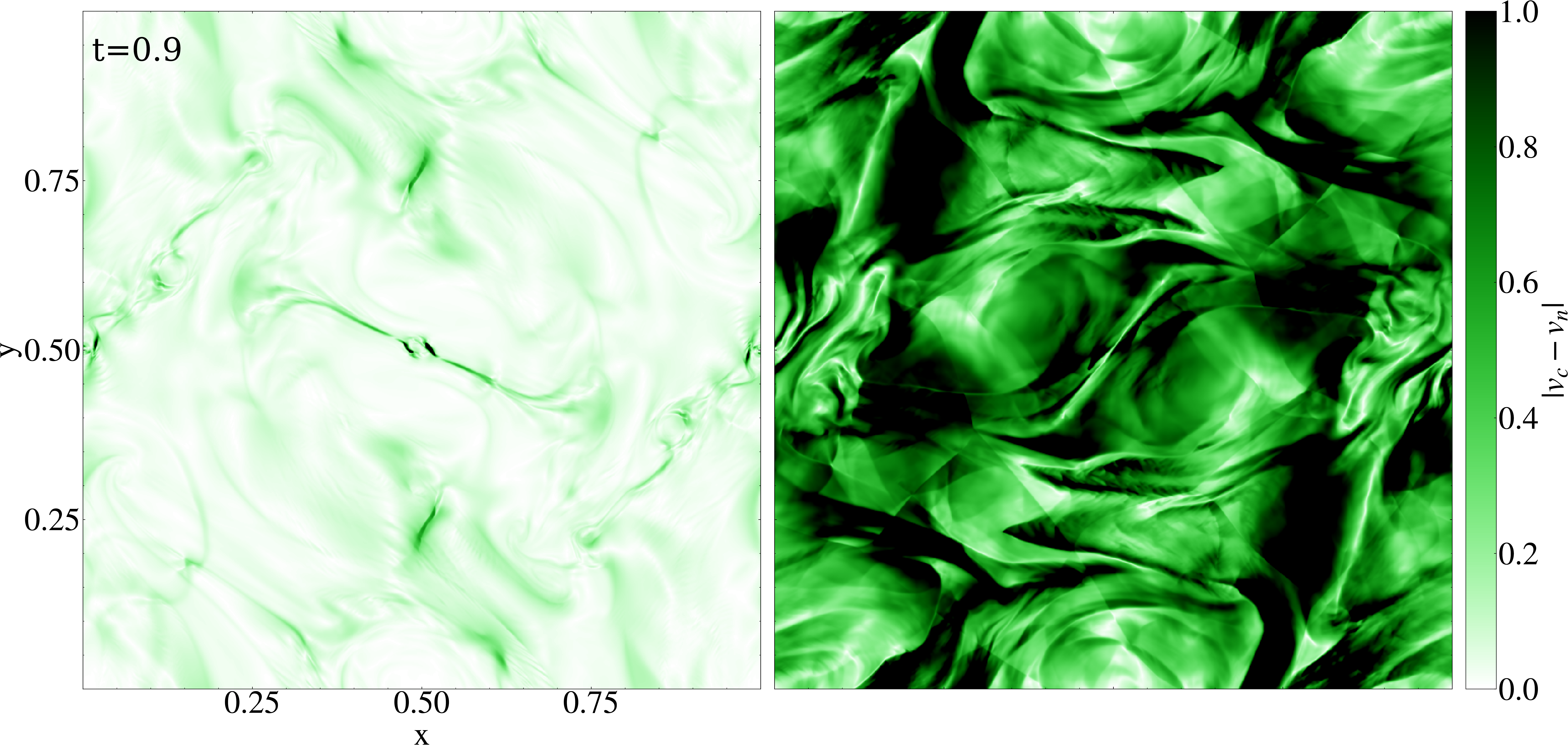}
}
\caption{
Magnitude of the decoupling in velocity between charges and neutrals.
Left: $\alpha=10^3$, Right: $\alpha=1$.
The colormap and the normalization are the same for both images.
}
\label{fig:ors3}
\end{figure}

On the other hand, in a weak coupling regime where we took $\alpha=1$ (now the mean free path becomes $\lambda_{in}=\lambda_{ni}=5.28$, larger than the length of the domain, though
of the same order of magnitude), the neutrals and charges evolve 
very differently and the density maps for both species are also very different than in the pure MHD case.
This can be seen in Figure \ref{fig:ors2}, where the densities of charges and neutrals are shown for the same moments of time as for the previous case.
The magnitude of the decoupling in velocity between neutrals and charges is much smaller in the strong coupling regime ($\alpha=10^3$)
than in the weak coupling regime ($\alpha=1$), as expected and quantified in Figure \ref{fig:ors3}.  


\begin{figure*}[!htb]
\centering
\FIG{
\includegraphics[width=8cm,height=9cm]{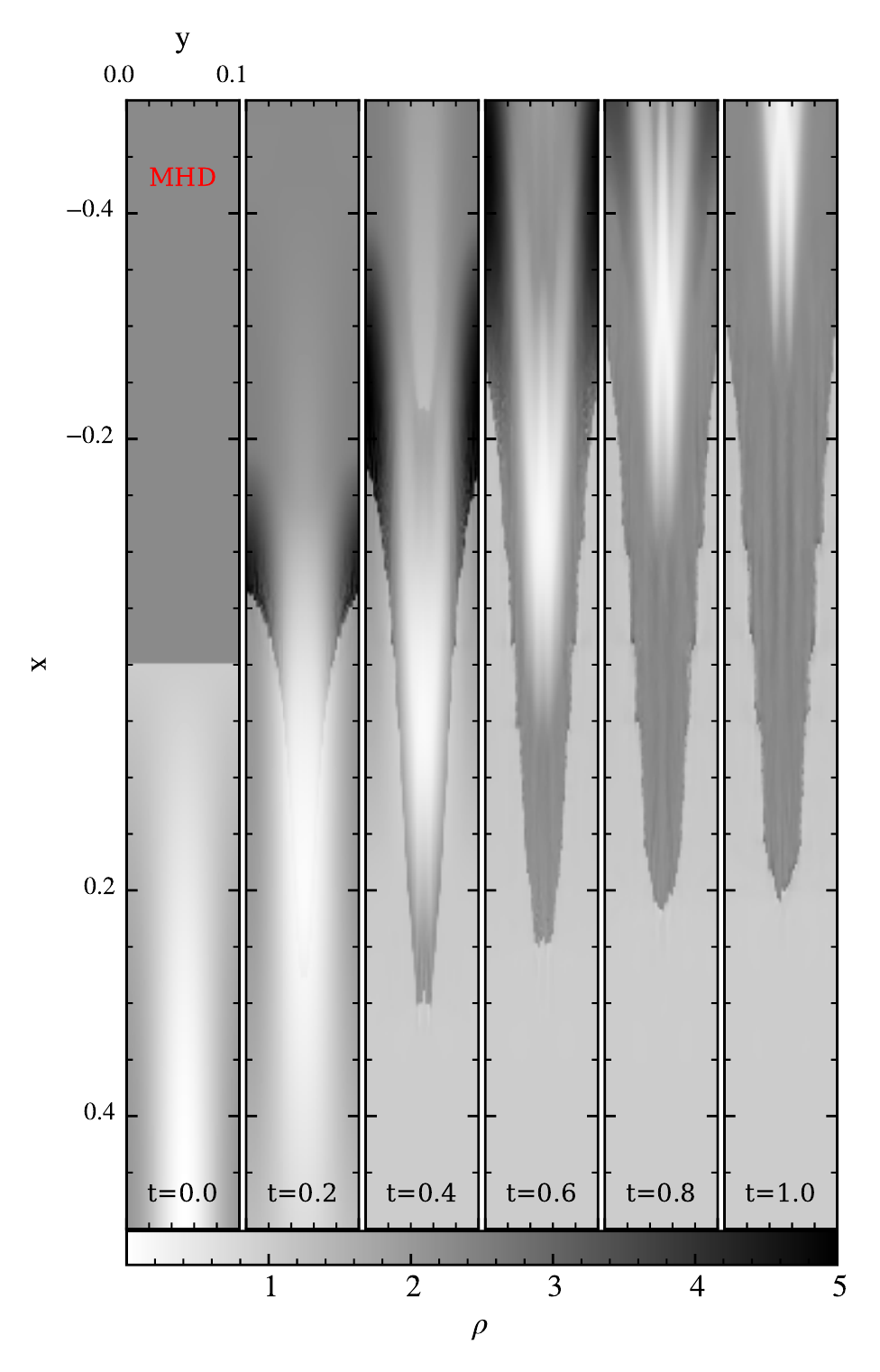}
\includegraphics[width=8cm,height=9cm]{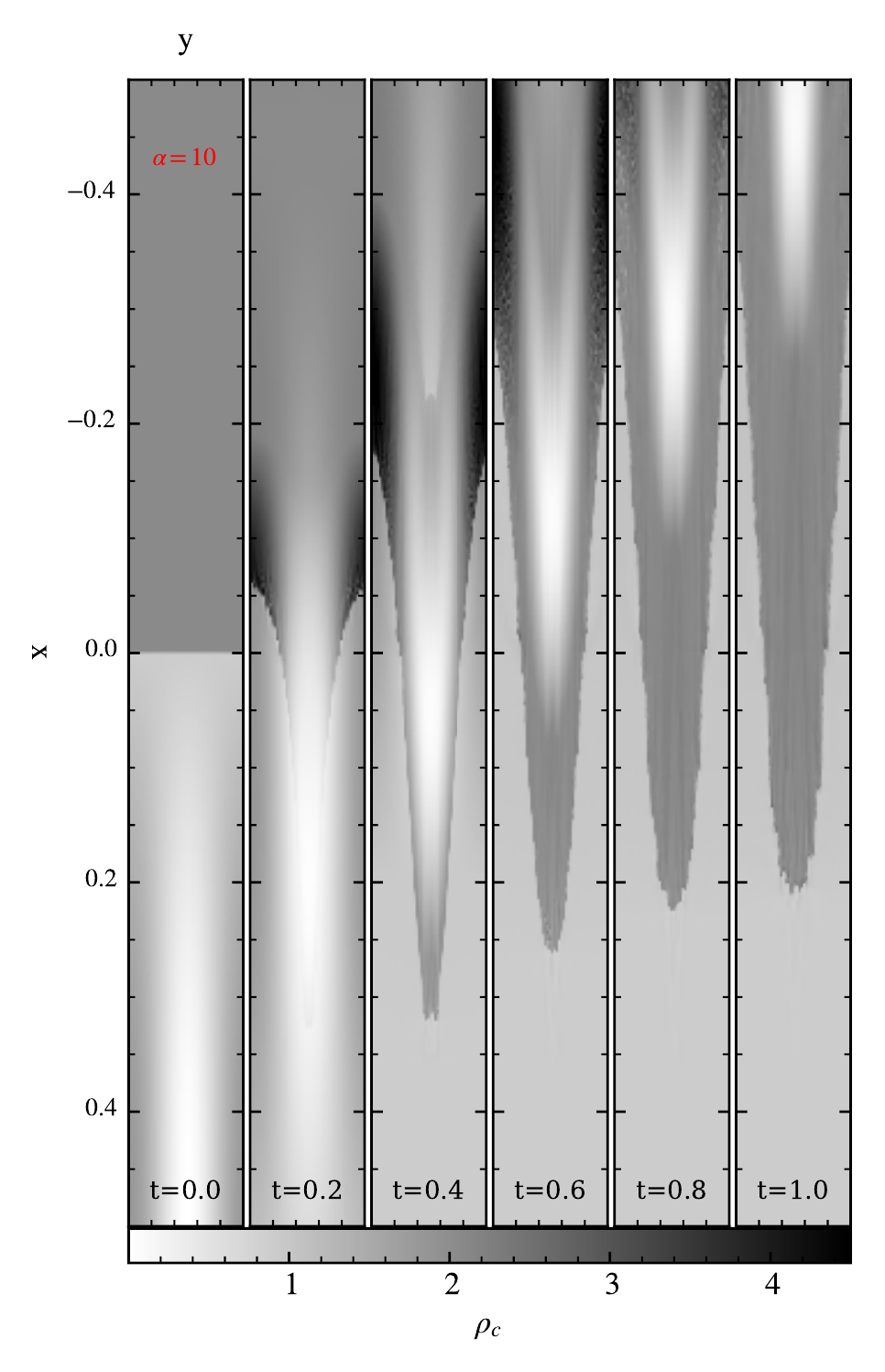}

\includegraphics[width=8cm,height=9cm]{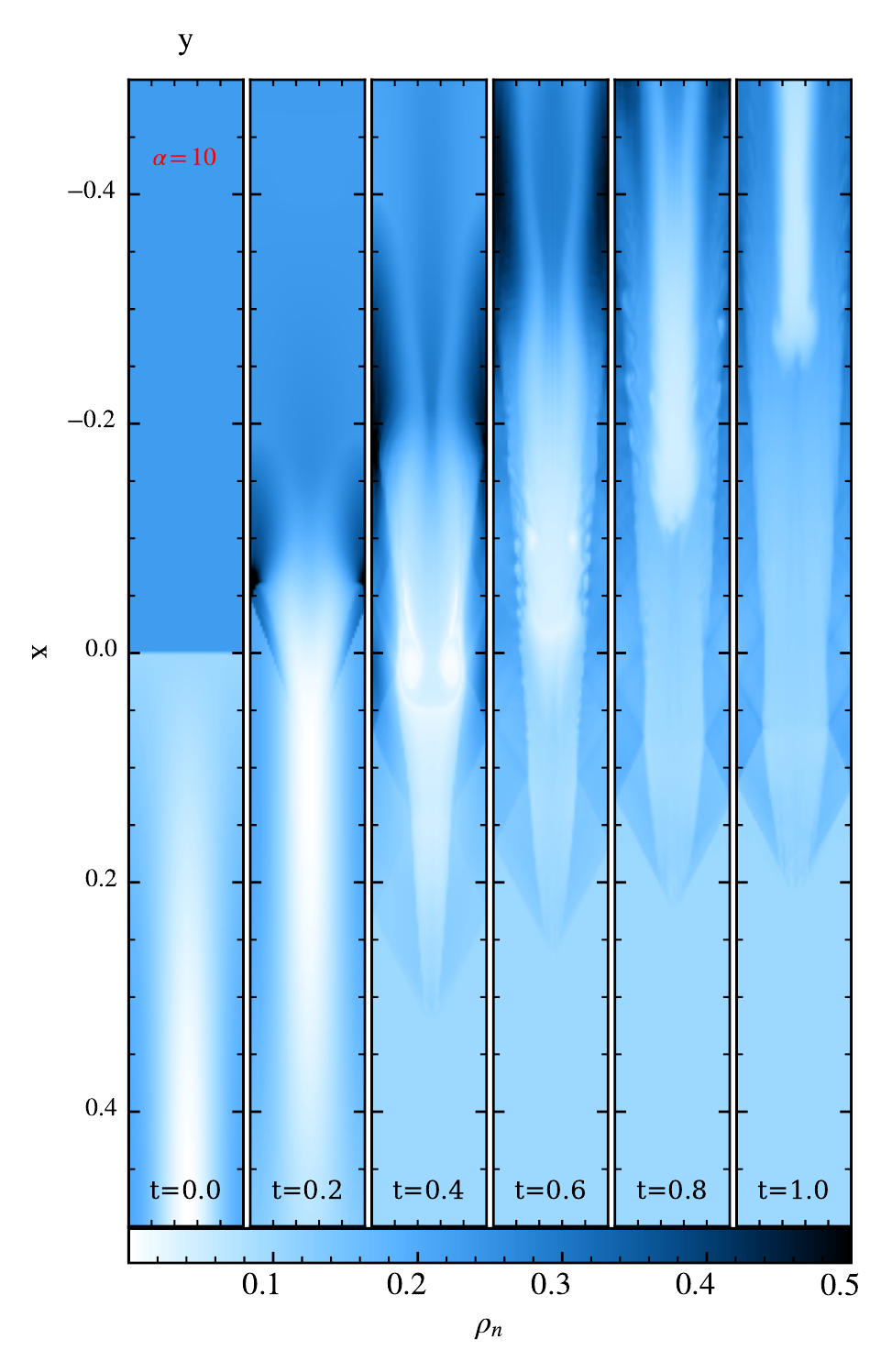}
\includegraphics[width=8cm,height=9cm]{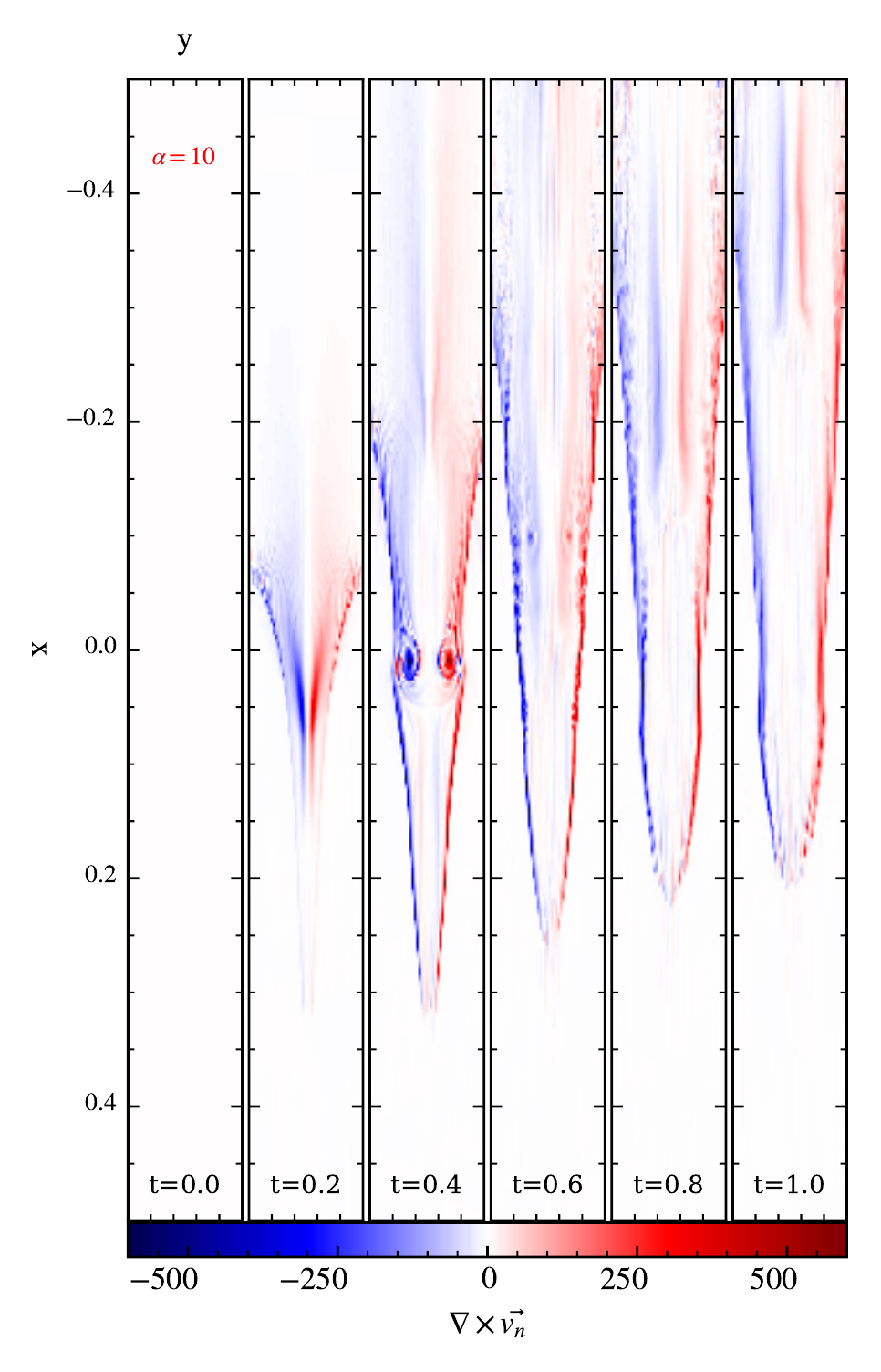}
}
\caption{
Time sequence for quantities seen in parallel shocks undergoing corrugation instabilities.  
Top left: Density in the MHD simulation. Top right: Density of charges in the two-fluid simulation.
Bottom left: Density of neutrals in the two-fluid simulation.
Bottom right: (Out-of-plane component of) Vorticity of neutrals in the two-fluid simulation.
The two-fluid simulation used a value of $\alpha=10$.
}
\label{fig:corru}
\end{figure*}

\subsection{Corrugation instability}

Differently to the Orszag-Tang test presented earlier, 
the charged fluid might evolve similarly to the plasma in the MHD assumption, even if the collisional coupling is weak. 
A very demanding numerical experiment is the simulation of a perturbed 2D MHD slow shock front, leading 
in most of the cases to the corrugation instability \citep{corr-stone,corr-ben}. 
We use a background medium similar to that used by \cite{corr-ben} for parallel shocks.
We use subscript $1$ for downstream variables ($x<0$) and subscript for $2$ upstream variables ($x>0$), where we exploit 
the frame of reference of the shock front, fixed at the location $x=0$. 
Given upstream quantities, we can get downstream quantities from the flux conservation laws, assuming a stationary solution.
We use the same values given in Table~A1 in \cite{corr-ben} for the left and right states with respect to interface of discontinuity.
The 2D $x-y$ domain $[-1.5,1] \times [0,0.1]$ is covered by $1024\times 256$ points as base resolution and we used 4 levels of refinement. 
The numerical values for the quantities upstream ($_2$) and downstream ($_1$)  are \citep{corr-ben},
\begin{eqnarray}
\left( \rho, {v_x}, {v_y}, p, {B_x}, {B_y} \right)_2 = \left( 1,-2,0,0.6,3.464, 0 \right)\,,\nonumber\\ 
\left( \rho, {v_x}, {v_y}, p, {B_x}, {B_y} \right)_1 = \left( 2.286,-0.875,0,2.85,3.464,0 \right)\,,
\end{eqnarray}

This planar shock is perturbed by its encounter with a controlled density perturbation. 
Therefore in the initial condition, the density upstream becomes $\rho_2+\delta\rho(x,y)$. 
This perturbation in the density is located upstream, only in the region $0\le x \le 1$:
\begin{equation}
\delta \rho = A\, \text{sin}\left(\pi x \right)  \text{cos}\left(\frac{2\pi y}{0.1}\right)\,.
\end{equation}

We use the same amplitude $A=1$ as \cite{corr-ben}, so that the simulation enters the non-linear regime from the beginning,
while \cite{corr-stone} used the value $A=10^{-4}$ in order to study the linear phase.
In order to avoid non-linear interaction between 
different modes, we used a single-mode corresponding to one wavelength.
The pure MHD result from \cite{corr-stone} shows that the instability grows faster for smaller modes.
As we used a smaller domain, compared to \cite{corr-ben}, we expect the instability to grow faster than in their case.

The two-fluid setup is obtained from the MHD setup using Eqs.~(\ref{eq:mhd-2fl}) and  $\xi_i=0.9$.
The value of the collisional parameter is uniform and constant, $\alpha=10$. 
The mean free path between ions and neutrals, calculated  according to Eq.~(\ref{eq:lambda}), $\lambda_{in} \approx 3.6$, 
is larger than the mean free path between neutrals and ions, $\lambda_{ni} \approx 0.4$,
both being larger than the size of the 
domain (in the direction $y$, which is horizontal in Figure~\ref{fig:corru}), meaning a weakly coupled regime. 

Figure~\ref{fig:corru} shows the numerical results of the fluid density obtained in the MHD assumption (top left panel)
and the charged fluid density in the two-fluid approximation (top right panel).
Bottom panels show the density (bottom left panel) and the vorticity of the neutrals (bottom right panel).
We observe that the charged fluid in the two-fluid simulation  evolves similarly to the MHD case, although some of the structures
are smoothed out by collisions.
In the neutral fluid (right panels), we notice that secondary Kelvin-Helmholtz instabilities (KHI) form as the perturbation advances towards the negative region of the $x$-domain. 
For this reason we plotted the vorticity of the neutrals, and indeed we see non-zero vorticity consistently with the locations where the vortices in the density are observed. 
Similarly to \cite{corr-ben}, the neutrals seem to stabilize the corrugation instability and present features as shock channels.
Differently from the simulations of \cite{corr-ben}, we used smaller wavelength of the perturbation, which grows faster, therefore
we analyzed the snapshots at an earlier time, while the perturbation was still travelling through the physical domain, producing KHI.

As smaller scales grow faster, we do see
small scale corrugation instabilities due to numerical noise appear along the deformed shock front. 
These secondary instabilities are not seen in the original two-fluid simulations of parallel shocks of \cite{corr-ben}.
{As we used very high resolution and no artificial dissipation, we claim these are physically relevant and further consequences of these finest-scale structures could become a subject of further study. }

\begin{figure*}[!htb]
\centering
\FIG{
\includegraphics[width=16cm]{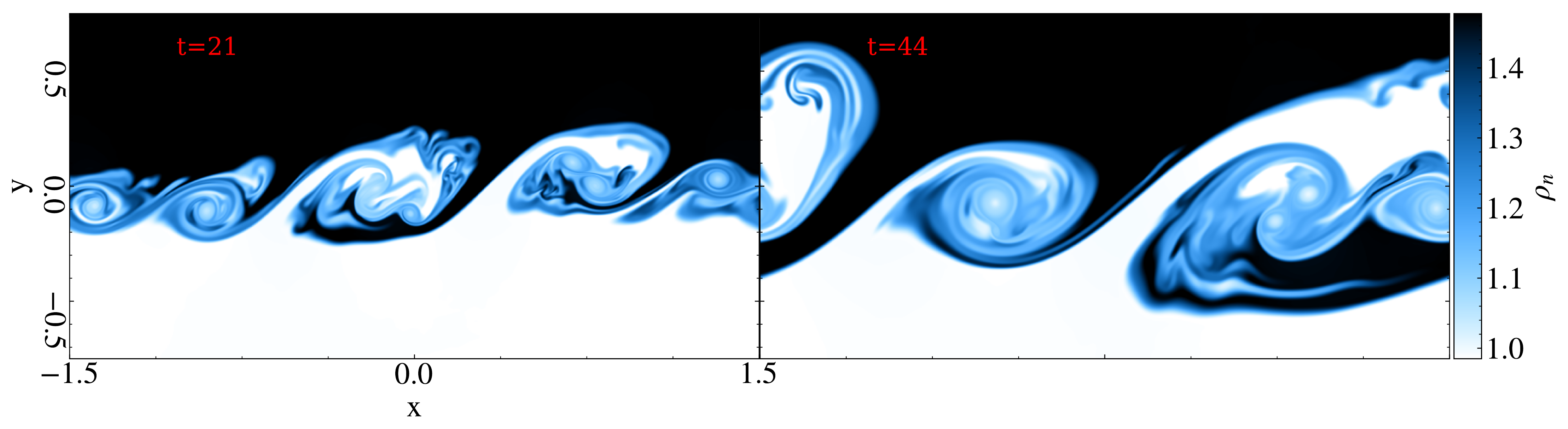}
}
\caption{
Time evolution of KHI-unstable neutral density in the uncoupled case ($\alpha=0$). 
The flux limiter used is ``cada3''. 
}
\label{fig:khi1}
\end{figure*}
\begin{figure*}[!htb]
\centering
\FIG{
\includegraphics[width=16cm]{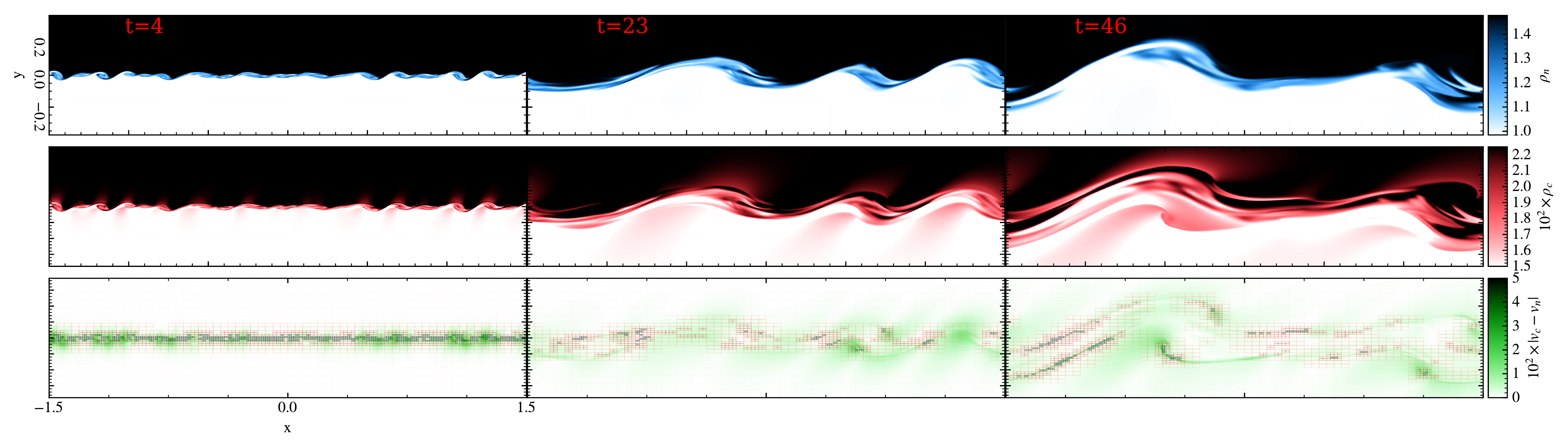}
}
\caption{
Time evolution for the coupled case. Top row: neutral density; Middle row: charged density; Bottom row: magnitude of the decoupling
in velocity between neutrals and charges. The AMR grid is overplotted in the bottom row plots. The vertical domain is between
-0.375 and 0.375, corresponding to the inner half of the full vertical domain.
The flux limiter used is ``cada3''. 
}
\label{fig:khi2}
\end{figure*}
\begin{figure*}[!htb]
\centering
\FIG{
\includegraphics[width=16cm]{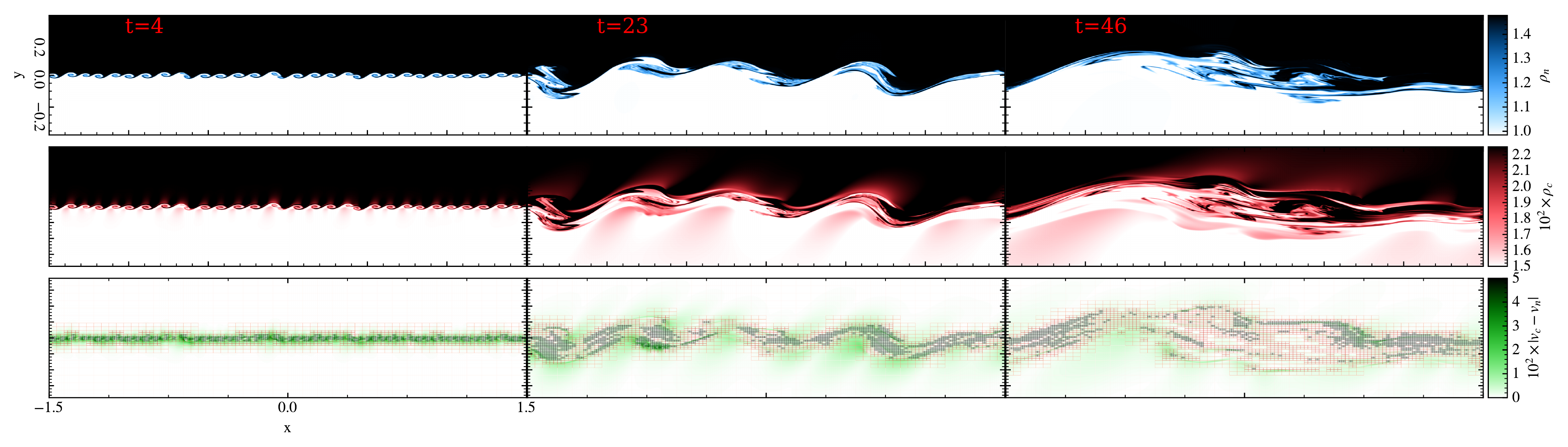}
}
\caption{
Same as Figure \ref{fig:khi2} for ``mp5'' limiter. 
}
\label{fig:khi3}
\end{figure*}
\begin{figure}[!h]
\centering
\FIG{
\includegraphics[width=8cm]{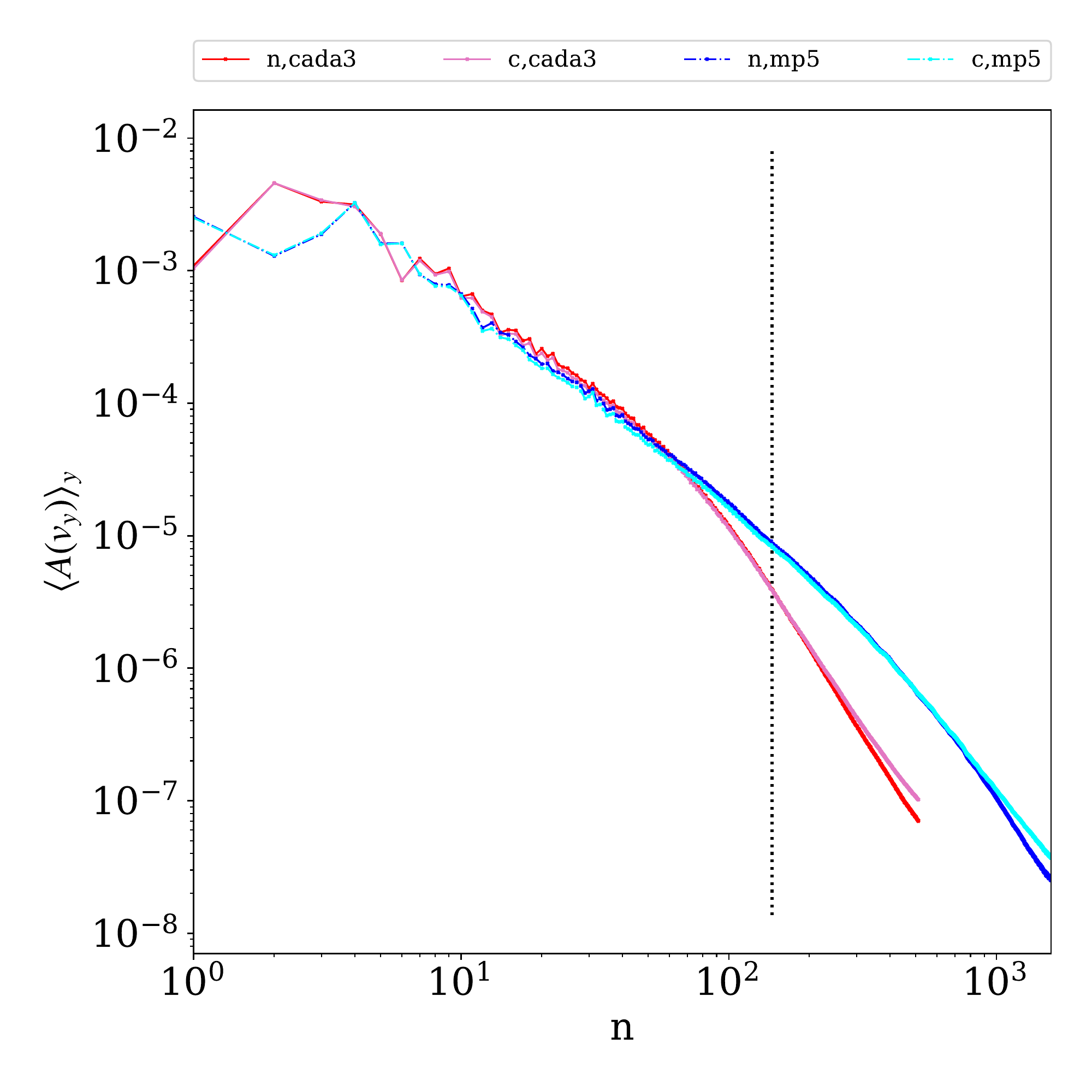}
}
\caption{
Fourier amplitudes of the $y$-velocities. The amplitudes are averaged in height between -0.375 and 0.375, the same height shown in Figures \ref{fig:khi2} and \ref{fig:khi3}.
The vertical dotted black line located at $n=145$ corresponds to the mode corresponding to the collision frequency between neutrals and ions, calculated from Eq. \ref{eq:col_freq}.
For a better visualization the upper limit on the $x$-axis is 512 for ``cada3'' limiter, and 1600 for  ``mp5'' limiter.
Both limiters show decoupling at small scales in the velocity.
}
\label{fig:khi4}
\end{figure}

\subsection{Two-fluid Kelvin-Helmholtz evolutions}

As stated above, the evolution of the parallel shock corrugation showed KHI in the neutrals. Here, we now turn to pure KHI studies, as done recently in a two-fluid setting.
We do simulations of 2D KHI using the same setup as \cite{khi-andrew}.
The domain is comprised between $-1.5$ and $1.5$ in the horizontal (aligned with the main shear flow) direction and between $-0.75$ and $0.75$ 
in the vertical direction.
We used AMR with a base resolution of $2048\times1024$ with 4 levels of refinement, thereby reaching the
same effective resolution as the (extremely high resolution) simulations of \cite{khi-andrew}.
We used the IMEX-ARS3 as temporal scheme and the HLL method for the calculation of the fluxes \citep{hll}.
The condition $\nabla \cdot \mathbf{B}=0$ is ensured by the parabolic cleaning method \citep{linde,tothdivb} specified by the value ``linde'' for the parameter typedivbfix. This method
adds non-conservative source terms in the induction and charged fluid energy equations, which diffuses the divergence of the magnetic field error at a maximal rate. 

The background in the MHD approximation is: 
\begin{eqnarray}
\label{eq:khi-ini}
\rho_1=1\,,\rho_2 =1.5\,,\nonumber\\
v_{\rm x1}=-\frac{1.5}{2.5} \Delta V \,,v_{\rm x2}=\frac{1}{2.5} \Delta V\,,
\text{where } \Delta V=0.2 \,,\nonumber\\
\beta=2\times 10^2\,,p=\frac{1}{\gamma}\,.
\end{eqnarray}
The subscripts $1$ and $2$ here represent quantities below ($y<0$) and above ($y>0$) the {sharp shear flow} interface located at $y=0$.
The initial magnetic field {(like the pressure $p$) is uniform and} has only a horizontal component $B_x = \sqrt{2p/\beta}$.
We then perturb {the planar interface with a small} $y$-velocity, being composed of 32 modes with random amplitude in $[-A,A]$ and phase in $[-\frac{\pi}{2} \frac{\pi}{2}]$, where 
\begin{equation}
A= 10^{-3} \sqrt{\gamma \frac{p}{\rho_1}} \,.
\end{equation}
This represents a very subsonic and linear, but still somewhat randomized, perturbation. 
This MHD-like initial condition is then transformed to a two-fluid initial condition by using Eqs.~(\ref{eq:mhd-2fl}) and
the ionization fraction $\xi_i = 10^{-2}$.

The limit of infinite collisional scale is the case when the collisional parameter $\alpha=0$.
In this fully uncoupled case, only the neutrals will evolve as for the charges the scales considered are smaller than the scale where the tension by the magnetic field stabilizes the KHI. Indeed, in the ideal MHD assumption, the horizontal magnetic field suppresses the KHI \citep[see Eqns. (205) or (15) in][respectively]{Ch1961, khi-andrew} if:  
\begin{equation}
\label{eq:khi_c}
(v_{\rm x1}-v_{\rm x2})^2 \le \frac{2 B_x^2 (\rho_1 + \rho_2) }{\rho_1 \rho_2}\,.
\end{equation}
Using the values from Eq.~(\ref{eq:khi-ini}) in the above  Eq. (\ref{eq:khi_c}), 
the value of the right hand side is of the same order of magnitude, but slightly smaller than the value of the left hand side
and this condition is not fulfilled, so the MHD case is unstable. 
In the uncoupled case, however, the density of interest should be replaced by the value of the charged fluid density only,
and for the very small ionization fraction considered here, the value of the right hand side is then almost two orders of magnitude
larger than left hand side. Hence, the magnetic field suppresses the instability in the charged fluid. 

Figure \ref{fig:khi1} shows two snapshots of the neutral density in the non-linear stage of the instability, in the fully uncoupled case.
During the non-linear stage of the instability large vortices form because of the inverse cascade, but there is also
a direct cascade when smaller scales form mainly due to secondary KHI and 
{(centrifugally induced) Rayleigh-Taylor effects}, similar to those seen in Figure 1 from \cite{khi-andrew}. 
However, we do not observe such small  scale structures as in the simulations of \cite{khi-andrew}. One of the reasons might be the fact that our initial perturbation
contains only 32 modes, differently to the white noise perturbation used by \cite{khi-andrew}.
However, the detailed evolution of the simulation in the nonlinear phase might be impacted by our choice of the flux limiter: this uncoupled case used a third order limiter \citep{cada3} (specified by the keyword ``cada3'' in {\tt MPI-AMRVAC}). 

In order to better understand the impact of the flux limiter, we show results obtained with two different limiters when we study the case with a finite mean free path between charges and neutrals.
For this coupled case, we use the value $\alpha=300$ as also presented in \cite{khi-andrew}. 
As we have seen in previous sections, for cases which had similar hydrodynamic scales, this value corresponds to a strong coupling regime.
Figures \ref{fig:khi2} and \ref{fig:khi3} show the density evolution of neutrals and charges, and the magnitude of the decoupling
in velocity between neutrals and charges for third order limiter ``cada3''
and a fifth order limiter \citep{mp5}, specified by the keyword ``mp5'' in {\tt MPI-AMRVAC}. 

In our AMR runs, the magnitude of the decoupling was used for the refinement criterion and we overplot the grids as well.
We observe how the finer grids nicely follow the regions with higher values of the decoupling. 
The third order limiter has more numerical diffusivity associated, therefore there is less detail in the density snapshots
corresponding to this simulation. As small scales are influenced by numerical diffusivity, the nonlinear process of merging of smaller scales to
larger scales happens faster for the ``cada3'' limiter than for ``mp5''. This can clearly be seen in the earliest snapshots corresponding to $t=4$.

The main conclusion of \cite{khi-andrew} was that at small scales neutrals and charges are decoupled while they are coupled at large scales.
In order to quantify the degree of coupling between charges and neutrals for different scales  we plot the Fourier amplitudes of the $y$-velocity. 
Figure \ref{fig:khi4} shows the Fourier amplitudes of the $y$-velocity for the two limiters ``cada3'' and ``mp5''.
The amplitudes are averaged in time
between $t=4$ and $t=46$ and also averaged in the vertical direction, restricted to the region $\{ y\vert -0.375 \le  y\le 0.375  \}$ shown in the figures. The Fourier modes
are shown as a function of the mode number $n = \frac{k L_x}{2 \pi}$, where $k$ is the wave number.

For {the adopted} low ionization fraction, the mean free path between neutrals and charges is larger
than the mean free path between charges and neutrals  and we calculate the mode number associated to this scale as:
\begin{equation}
\label{eq:col_freq}
{n_{\rm coll}}_{ni} = \frac{L_x}{2 \pi} \frac{1}{\lambda_{ni}}
\end{equation}
where $\lambda_{ni}$ has been defined in Eq.~(\ref{eq:lambda}).
The value obtained ${n_{\rm coll}}_{ni}=145$ is shown by a dotted black line, and ${n_{\rm coll}}_{in}=99 \times {n_{\rm coll}}_{ni}$ does not appear in the plot.
We can see in Figure~\ref{fig:khi4} that the curves corresponding to the neutral and charged fluids overlap for large scales, but diverge for small scales.
For the ``mp5'' limiter, which has smaller numerical diffusivity associated, the decoupling appears at  smaller scales than for ``cada3''.
The largest scales at which the decoupling appears visually for both limiters are smaller than the estimated decoupling scale between neutrals and ions, but
larger than the estimated decoupling scale between ions and neutrals, so the result of the numerical simulations are consistent with this theoretical estimation.
Similarly to \cite{khi-andrew}, the velocity of charges and neutrals are coupled for large scales and decoupled on the scales
smaller than the mean free path between the particles of different species.
\section{Conclusions}
\label{sec:concl}
We implemented the full two-fluid (charges-neutrals) equations in {\tt MPI-AMRVAC}, an open-source, versatile code that can be used in many astrophysical contexts.
The implementation shares many aspects with the {\textsc{Mancha3D-2F}} code, described in \cite{Popescu+etal2018}, but the {\tt MPI-AMRVAC} code offers various advantages, such as being fully dimension-independent (1D, 2D or 3D), and having the option to use an efficient block-adaptive AMR mesh.
The use of an IMEX scheme, with implicit updates for the collisional terms,
greatly decreases the computational cost in highly collisional regimes.
The modular structure of the code permitted us to use already implemented numerical discretizations (both temporal and spatial) and parallelization algorithms in the two-fluid model, and exploit the AMR. 
The use of AMR further reduces the computational cost, compared to the uniform grid with the same effective resolution.
Many two-fluid codes, such as those in \citep{Popescu+etal2018,2017Kuzma,2016Hillier} use uniform grid implementations. 
We performed many tests of the code which showed physically sound results, and were consistent to other results described in recent literature. 
The conclusions are summarized below. 
\begin{itemize}
\item
The spatial and temporal convergence tests showed that the numerical solution converges to the analytical solution for simulations of
Alfv\'en waves in uniform settings.
The numerical error was smaller for smaller coupling parameter $\alpha$, similarly to results of \cite{Popescu+etal2018}. {The IMEX-ARS3 scheme, together with standard shock-capturing discretizations, achieves high-order (2nd to third order) convergence.}
\item
The simulations done with our newly implemented module in the {\tt MPI-AMRVAC} code reproduced and extended
previous results on waves, shocks and instabilities, first obtained with other codes.
\item 
When the collisional timescales are smaller than the hydrodynamical timescales (well coupled regime), the effect of the collisions is similar to diffusivity.
The collisions can damp waves, similarly to the conclusion of \cite{Diaz2012,Popescu2018b} and decrease the growth of the instabilities \citep{Popescu3}.
We observed damping of waves and decoupling in velocity in 1D (Figure~\ref{fig:maw}) and 2D (Figure~\ref{fig:w2f}) setups.
\item
The numerical results for the 1.75D Riemann problem (Figure~\ref{fig:shockc}) 
and the well-coupled Orszag-Tang (Figure~\ref{fig:ors1}) results in the two-fluid approach resembled results obtained in the MHD approach with diffusivity.
\item
For hydrodynamical timescales similar or smaller than the collisional timescales (weakly coupled regime), neutrals and charges might behave differently.
For the RTI studied by \cite{DiazKh2013} and \cite{2014bKh} the hydrodynamic scale 
defined by the scale of  magnetic cutoff was below the collisional timescale, 
therefore the neutrals, which do not feel the stabilizing effect of the magnetic field are unstable at these scales.
Our simulations that reproduced the simulations of KHI done by \cite{khi-andrew}, in absence of collisions, showed that the instability grows in neutrals
while the charged fluid did not evolve because of the magnetic field.
The simulations of Orszag-Tang test in weakly coupled regime (Figure~\ref{fig:ors2}) showed structures in the density of neutrals and charges
very different to the structures in density obtained in the MHD approximation.
However, simulations of the corrugation instability showed structures in the charged fluid similar to
the structures in the plasma obtained in the MHD approximation, slightly smoothed out (top panels of Figure~(\ref{fig:corru})).
The structures in the neutral fluid were different, and we could observe secondary KHI (bottom panels of Figure~\ref{fig:corru}). {The fact that we showed a two-fluid setup where neutrals undergo KHI while charges do not, illustrate the need for two-fluid modeling to find previously unexplored, interesting dynamics.}
\item 
Numerical diffusivity makes decoupling between charges and neutrals appear at larger scales (Figure~\ref{fig:khi4}).
When exploiting different limiters in the numerical reconstruction procedure, the simulations of KHI can be visually (slightly) different,
but statistical properties are similar.
The conclusion of \cite{khi-andrew}, that the charges and neutrals are decoupled at small scales, but coupled at large scales,
also applies for our simulations.  
\item
Two-fluid effects occur at small scales and the use of AMR permits very high resolution at a reasonable computational cost. 
\end{itemize}

Future work would include ionization/recombination effects and radiative losses
and address e.g. the secondary instabilities we observed in the corrugation instability study.
For simplicity, or in order to compare to analytical results or other existing results, the value of the collisional parameter
$\alpha$ was kept uniform and constant, however more realistic simulations should consider its dependence on local plasma parameters,
as described by Eq.~(\ref{eq:alpha_el}).
The viscosity and thermal conductivity might also have an important role in the study of waves and instabilities 
as a mechanism of dissipation and they might be similar to the two-fluid effects in highly collisional regime \citep{Popescu4}.
\begin{acknowledgements}
This work was supported by the FWO grant 1232122N and a FWO grant G0B4521N. 
This project has received funding from the European Research Council (ERC) under
the European Union’s Horizon 2020 research and innovation programme (grant
agreement No. 833251 PROMINENT ERC-ADG 2018). This research is further supported by Internal funds KU Leuven, through the project C14/19/089 TRACESpace. 
The resources and services used in this work were provided by the VSC (Flemish Supercomputer Center), funded by the Research Foundation - Flanders (FWO) and the Flemish Government.
\end{acknowledgements}

\bibliographystyle{aa}

\begin{thebibliography}{62}
\expandafter\ifx\csname natexlab\endcsname\relax\def\natexlab#1{#1}\fi

\bibitem[{{Anan} {et~al.}(2017){Anan}, {Ichimoto}, \& {Hillier}}]{2017Anan}
{Anan}, T., {Ichimoto}, K., \& {Hillier}, A. 2017, \aap, 601, A103

\bibitem[{Arber {et~al.}(2007)Arber, Haynes, \& Leake}]{Arber_2007}
Arber, T.~D., Haynes, M., \& Leake, J.~E. 2007, The Astrophysical Journal, 666,
  541

\bibitem[{Ascher {et~al.}(1997)Ascher, Ruuth, \& Spiteri}]{ARS3}
Ascher, U.~M., Ruuth, S.~J., \& Spiteri, R.~J. 1997, Applied Numerical
  Mathematics, 25, 151, special Issue on Time Integration

\bibitem[{{Ballester} {et~al.}(2018){Ballester}, {Carbonell}, {Soler}, \&
  {Terradas}}]{2018Ballester2}
{Ballester}, J.~L., {Carbonell}, M., {Soler}, R., \& {Terradas}, J. 2018, \aap,
  609, A6

\bibitem[{Chandrasekhar(1961)}]{Ch1961}
Chandrasekhar, S. 1961, Hydrodynamic and Hydromagnetic Stability, International
  series of monographs on physics (Clarendon Press)

\bibitem[{{Cui} \& {Bai}(2021)}]{ad-protopl2}
{Cui}, C. \& {Bai}, X.-N. 2021, \mnras, 507, 1106

\bibitem[{{de la Cruz Rodr{\'{\i}}guez} \&
  {Socas-Navarro}(2011)}]{2011delaCruz}
{de la Cruz Rodr{\'{\i}}guez}, J. \& {Socas-Navarro}, H. 2011, \aap, 527, L8

\bibitem[{{D\'{\i}az} {et~al.}(2014){D\'{\i}az}, {Khomenko}, \&
  {Collados}}]{DiazKh2013}
{D\'{\i}az}, A.~J., {Khomenko}, E., \& {Collados}, M. 2014, A\&A, 564, A97

\bibitem[{D{\'{\i}}az {et~al.}(2012)D{\'{\i}}az, Soler, \&
  Ballester}]{Diaz2012}
D{\'{\i}}az, A.~J., Soler, R., \& Ballester, J.~L. 2012, The Astrophysical
  Journal, 754, 41

\bibitem[{{Draine} \& {McKee}(1993)}]{Draine1993}
{Draine}, B.~T. \& {McKee}, C.~F. 1993, \araa, 31, 373

\bibitem[{{Draine} {et~al.}(1983){Draine}, {Roberge}, \&
  {Dalgarno}}]{1983Draine}
{Draine}, B.~T., {Roberge}, W.~G., \& {Dalgarno}, A. 1983, \apj, 264, 485

\bibitem[{{Felipe} {et~al.}(2010){Felipe}, {Khomenko}, \&
  {Collados}}]{Felipe2010}
{Felipe}, T., {Khomenko}, E., \& {Collados}, M. 2010, \apj, 719, 357

\bibitem[{{Gilbert} {et~al.}(2007){Gilbert}, {Kilper}, \&
  {Alexander}}]{2007Gilbert}
{Gilbert}, H., {Kilper}, G., \& {Alexander}, D. 2007, \apj, 671, 978

\bibitem[{{Gilbert} {et~al.}(2002){Gilbert}, {Hansteen}, \&
  {Holzer}}]{2002Gilbert}
{Gilbert}, H.~R., {Hansteen}, V.~H., \& {Holzer}, T.~E. 2002, \apj, 577, 464

\bibitem[{Goedbloed {et~al.}(2019)Goedbloed, Keppens, \& Poedts}]{hans2019}
Goedbloed, H., Keppens, R., \& Poedts, S. 2019, Magnetohydrodynamics of
  Laboratory and Astrophysical Plasmas (Cambridge University Press)

\bibitem[{{Hillier}(2019{\natexlab{a}})}]{ref1}
{Hillier}, A. 2019{\natexlab{a}}, arXiv e-prints, arXiv:1907.12507

\bibitem[{{Hillier}(2019{\natexlab{b}})}]{khi-andrew}
{Hillier}, A. 2019{\natexlab{b}}, Physics of Plasmas, 26, 082902

\bibitem[{{Hillier} {et~al.}(2016){Hillier}, {Takasao}, \&
  {Nakamura}}]{2016Hillier}
{Hillier}, A., {Takasao}, S., \& {Nakamura}, N. 2016, \aap, 591, A112

\bibitem[{{Keppens} {et~al.}(2003){Keppens}, {Nool}, {T{\'o}th}, \&
  {Goedbloed}}]{tothdivb}
{Keppens}, R., {Nool}, M., {T{\'o}th}, G., \& {Goedbloed}, J.~P. 2003, Computer
  Physics Communications, 153, 317

\bibitem[{{Khomenko} \& {Cally}(2019)}]{ambi2}
{Khomenko}, E. \& {Cally}, P.~S. 2019, \apj, 883, 179

\bibitem[{{Khomenko} {et~al.}(2014{\natexlab{a}}){Khomenko}, {Collados},
  {D{\'\i}az}, \& {Vitas}}]{eqkh}
{Khomenko}, E., {Collados}, M., {D{\'\i}az}, A., \& {Vitas}, N.
  2014{\natexlab{a}}, Physics of Plasmas, 21, 092901

\bibitem[{{Khomenko} {et~al.}(2016){Khomenko}, {Collados}, \&
  {D{\'{\i}}az}}]{2016Khomenko}
{Khomenko}, E., {Collados}, M., \& {D{\'{\i}}az}, A.~J. 2016, \apj, 823, 132

\bibitem[{Khomenko {et~al.}(2015)Khomenko, Collados, Shchukina, \&
  D{\'{\i}}az}]{2015Khomenko}
Khomenko, E., Collados, M., Shchukina, N., \& D{\'{\i}}az, A. 2015, \aap, 584,
  A66

\bibitem[{{Khomenko} {et~al.}(2014{\natexlab{b}}){Khomenko}, {D{\'{\i}}az}, {de
  Vicente}, {Collados}, \& {Luna}}]{2014bKh}
{Khomenko}, E., {D{\'{\i}}az}, A., {de Vicente}, A., {Collados}, M., \& {Luna},
  M. 2014{\natexlab{b}}, \aap, 565, A45

\bibitem[{{Ku{\'z}ma} {et~al.}(2017){Ku{\'z}ma}, {Murawski}, {Kayshap},
  {W{\'o}jcik}, {Srivastava}, \& {Dwivedi}}]{2017Kuzma}
{Ku{\'z}ma}, B., {Murawski}, K., {Kayshap}, P., {et~al.} 2017, \apj, 849, 78

\bibitem[{{Lesur}(2020)}]{ad-protopl}
{Lesur}, G. 2020, arXiv e-prints, arXiv:2007.15967

\bibitem[{{MacNeice} {et~al.}(2000){MacNeice}, {Olson}, {Mobarry}, {de
  Fainchtein}, \& {Packer}}]{paramesh}
{MacNeice}, P., {Olson}, K.~M., {Mobarry}, C., {de Fainchtein}, R., \&
  {Packer}, C. 2000, Computer Physics Communications, 126, 330

\bibitem[{Marder(1987)}]{linde}
Marder, B. 1987, Journal of Computational Physics, 68, 48

\bibitem[{{Mart{\'{\i}}nez-G{\'o}mez}
  {et~al.}(2016){Mart{\'{\i}}nez-G{\'o}mez}, {Soler}, \&
  {Terradas}}]{2016Gomez1}
{Mart{\'{\i}}nez-G{\'o}mez}, D., {Soler}, R., \& {Terradas}, J. 2016, \apj,
  832, 101

\bibitem[{{Mart{\'{\i}}nez-G{\'o}mez}
  {et~al.}(2017){Mart{\'{\i}}nez-G{\'o}mez}, {Soler}, \&
  {Terradas}}]{2017Gomez1}
{Mart{\'{\i}}nez-G{\'o}mez}, D., {Soler}, R., \& {Terradas}, J. 2017, \apj,
  837, 80

\bibitem[{O'Flannagain {et~al.}(2015)O'Flannagain, Brown, \& Gallagher}]{ionFr}
O'Flannagain, A., Brown, J., \& Gallagher, P. 2015, The Astrophysical Journal,
  799, 127

\bibitem[{{Orszag} \& {Tang}(1979)}]{orszag1}
{Orszag}, S.~A. \& {Tang}, C.~M. 1979, Journal of Fluid Mechanics, 90, 129

\bibitem[{Picone \& Dahlburg(1991)}]{orszag2}
Picone, J.~M. \& Dahlburg, R.~B. 1991, Physics of Fluids B: Plasma Physics, 3,
  29

\bibitem[{{Popescu Braileanu} \& {Keppens}(2021)}]{beatrice}
{Popescu Braileanu}, B. \& {Keppens}, R. 2021, \aap, 653, A131

\bibitem[{{Popescu Braileanu} {et~al.}(in revision){Popescu Braileanu},
  {Lukin}, \& {Khomenko}}]{Popescu5}
{Popescu Braileanu}, B., {Lukin}, V.~S., \& {Khomenko}, E. in revision, arXiv
  e-prints, arXiv:2112.13043

\bibitem[{{Popescu Braileanu} {et~al.}(2019{\natexlab{a}}){Popescu Braileanu},
  {Lukin}, {Khomenko}, \& {de Vicente}}]{Popescu+etal2018}
{Popescu Braileanu}, B., {Lukin}, V.~S., {Khomenko}, E., \& {de Vicente},
  {\'A}. 2019{\natexlab{a}}, \aap, 627, A25

\bibitem[{{Popescu Braileanu} {et~al.}(2019{\natexlab{b}}){Popescu Braileanu},
  {Lukin}, {Khomenko}, \& {de Vicente}}]{Popescu2018b}
{Popescu Braileanu}, B., {Lukin}, V.~S., {Khomenko}, E., \& {de Vicente},
  {\'A}. 2019{\natexlab{b}}, \aap, 630, A79

\bibitem[{{Popescu Braileanu} {et~al.}(2021{\natexlab{a}}){Popescu Braileanu},
  {Lukin}, {Khomenko}, \& {de Vicente}}]{Popescu3}
{Popescu Braileanu}, B., {Lukin}, V.~S., {Khomenko}, E., \& {de Vicente},
  {\'A}. 2021{\natexlab{a}}, \aap, 646, A93

\bibitem[{{Popescu Braileanu} {et~al.}(2021{\natexlab{b}}){Popescu Braileanu},
  {Lukin}, {Khomenko}, \& {de Vicente}}]{Popescu4}
{Popescu Braileanu}, B., {Lukin}, V.~S., {Khomenko}, E., \& {de Vicente},
  {\'A}. 2021{\natexlab{b}}, \aap, 650, A181

\bibitem[{Smirnov(2003)}]{2003Smirnov}
Smirnov, B.~M. 2003, Physics of Atoms and Ions (Springer-Verlag New York),
  XIII, 443

\bibitem[{Smith \& Sakai(2008)}]{2008Sakai}
Smith, P.~D. \& Sakai, J.~I. 2008, \aap, 486, 569

\bibitem[{{Snow} \& {Hillier}(2019)}]{2019Snow}
{Snow}, B. \& {Hillier}, A. 2019, arXiv e-prints, arXiv:1904.12518

\bibitem[{{Snow} \& {Hillier}(2021)}]{corr-ben}
{Snow}, B. \& {Hillier}, A. 2021, \mnras, 506, 1334

\bibitem[{{Soler} \& {Ballester}(2022)}]{solerFr}
{Soler}, R. \& {Ballester}, J.~L. 2022, Frontiers in Astronomy and Space
  Science, 9

\bibitem[{{Soler} {et~al.}(2013{\natexlab{a}}){Soler}, {Carbonell},
  {Ballester}, \& {Terradas}}]{2013Soler}
{Soler}, R., {Carbonell}, M., {Ballester}, J.~L., \& {Terradas}, J.
  2013{\natexlab{a}}, \apj, 767, 171

\bibitem[{{Soler} {et~al.}(2013{\natexlab{b}}){Soler}, {Diaz}, {Ballester}, \&
  {Goossens}}]{2013Soler2}
{Soler}, R., {Diaz}, A.~J., {Ballester}, J.~L., \& {Goossens}, M.
  2013{\natexlab{b}}, A\&A, 551, A86

\bibitem[{{Stone} \& {Edelman}(1995)}]{corr-stone}
{Stone}, J.~M. \& {Edelman}, M. 1995, \apj, 454, 182

\bibitem[{{Suresh} \& {Huynh}(1997)}]{mp5}
{Suresh}, A. \& {Huynh}, H.~T. 1997, Journal of Computational Physics, 136, 83

\bibitem[{{Teunissen} \& {Keppens}(2019)}]{multigrid}
{Teunissen}, J. \& {Keppens}, R. 2019, Computer Physics Communications, 245,
  106866

\bibitem[{Toro(1997)}]{hll}
Toro, E.~F. 1997, The HLL and HLLC Riemann Solvers (Berlin, Heidelberg:
  Springer Berlin Heidelberg), 293--311

\bibitem[{{Torrilhon}(2003)}]{torrh}
{Torrilhon}, M. 2003, Journal of Plasma Physics, 69, 253

\bibitem[{Tóth {et~al.}(2012)Tóth, van~der Holst, Sokolov, Zeeuw, Gombosi,
  Fang, Manchester, Meng, Najib, Powell, Stout, Glocer, Ma, \&
  Opher}]{2012Toth}
Tóth, G., van~der Holst, B., Sokolov, I.~V., {et~al.} 2012, Journal of
  Computational Physics, 231, 870 , special Issue: Computational Plasma
  PhysicsSpecial Issue: Computational Plasma Physics

\bibitem[{{Vernazza} {et~al.}(1981){Vernazza}, {Avrett}, \& {Loeser}}]{valc}
{Vernazza}, J.~E., {Avrett}, E.~H., \& {Loeser}, R. 1981, \apjs, 45, 635

\bibitem[{Voronov(1997)}]{1997Voronov}
Voronov, G. 1997, Atomic Data and Nuclear Data Tables, 65, 1

\bibitem[{{Wiehr} {et~al.}(2019){Wiehr}, {Stellmacher}, \&
  {Bianda}}]{2019Wiehr}
{Wiehr}, E., {Stellmacher}, G., \& {Bianda}, M. 2019, \apj, 873, 125

\bibitem[{{Wurster} {et~al.}(2022){Wurster}, {Bate}, {Price}, \&
  {Bonnell}}]{ad-mol}
{Wurster}, J., {Bate}, M.~R., {Price}, D.~J., \& {Bonnell}, I.~A. 2022, \mnras,
  511, 746

\bibitem[{{Xia} {et~al.}(2018){Xia}, {Teunissen}, {El Mellah}, {Chan{\'e}}, \&
  {Keppens}}]{2018Rony}
{Xia}, C., {Teunissen}, J., {El Mellah}, I., {Chan{\'e}}, E., \& {Keppens}, R.
  2018, \apjs, 234, 30

\bibitem[{{Yadav} {et~al.}(2022){Yadav}, {Keppens}, \& {Popescu
  Braileanu}}]{Nitin}
{Yadav}, N., {Keppens}, R., \& {Popescu Braileanu}, B. 2022, arXiv e-prints,
  arXiv:2201.09704

\bibitem[{{Zaqarashvili} {et~al.}(2011{\natexlab{a}}){Zaqarashvili},
  {Khodachenko}, \& {Rucker, H. O.}}]{2011Zaq2}
{Zaqarashvili}, T.~V., {Khodachenko}, M.~L., \& {Rucker, H. O.}
  2011{\natexlab{a}}, A\&A, 534, A93

\bibitem[{{Zaqarashvili} {et~al.}(2011{\natexlab{b}}){Zaqarashvili},
  {Khodachenko}, \& {Rucker, H. O.}}]{2011Zaq}
{Zaqarashvili}, T.~V., {Khodachenko}, M.~L., \& {Rucker, H. O.}
  2011{\natexlab{b}}, A\&A, 529, A82

\bibitem[{{Zaqarashvili} {et~al.}(2013){Zaqarashvili}, {Khodachenko}, \&
  {Soler}}]{2013Zaq}
{Zaqarashvili}, T.~V., {Khodachenko}, M.~L., \& {Soler}, R. 2013, \aap, 549,
  A113

\bibitem[{Čada \& Torrilhon(2009)}]{cada3}
Čada, M. \& Torrilhon, M. 2009, Journal of Computational Physics, 228, 4118

\end{thebibliography}

\begin{appendix} 
\section{Expressions for the collisional terms} 
\label{app:collTerms}
The implementation is the same as done in {\textsc{Mancha3D-2F}} code \citep{Popescu+etal2018} which uses the SI unit system. 
\begin{equation} \label{eq:nu_el} 
 \nu_{\alpha\beta}= n_\beta \frac{m_\beta}{m_\alpha + m_\beta } \sqrt{\frac{8 k_B T_{\alpha\beta}}{\pi m_{\alpha\beta } }} \Sigma_{\alpha\beta} 
.\end{equation}
Here, $T_{\alpha\beta}= (T_\alpha +T_\beta)/2$ is the average temperature,  and $m_{\alpha\beta } = m_\alpha m_\beta/(m_\alpha + m_\beta ) $ is the reduced mass of particles $\alpha$ and $\beta$, {and $\Sigma_{\alpha\beta}$ is the corresponding collisional cross-section.}

The general expression for $\alpha$ is then the same as Eq. (A.4) from \cite{Popescu+etal2018}, and reads
\begin{equation} \label{eq:alpha_el_gen} 
\alpha = \frac{m_{in}}{{m_n}^2} \sqrt{ \frac{8 k_B T_{cn}}{\pi m_{in}}}  \Sigma_{in} + \frac{m_{en}}{{m_n}^2} \sqrt{ \frac{8 k_B T_{cn}}{\pi m_{en}}}  \Sigma_{en}
\,,\end{equation}
where $m_n=m_H$ is the Hydrogen mass. 
When the collisions with electrons  are neglected and considering $m_i = m_n$, the above Eq.~(\ref{eq:alpha_el_gen}) becomes: 
\begin{equation} \label{eq:alpha_el} 
\alpha = \frac{2}{{m_H}^{3/2} \sqrt{\pi}} \sqrt{ k_B T_{cn}}  \Sigma_{in} 
,\end{equation}
where the collisional cross-section considered here is $\Sigma_{in} = 10^{-19} \text{m}^{2}$. {This is the (dimensional) form for $\alpha$ used in all our simulations for this paper.}

Expressions for $\Gamma^{\rm ion}$ and $\Gamma^{\rm rec} $ as functions of $n_e$ and $T_e$ are given in \cite{1997Voronov} and \cite{2003Smirnov}
and are the same as Eqs. (A.2), (A.1) from \cite{Popescu+etal2018}:
\begin{equation} \label{eq:gamma_rec}
\Gamma^{\rm rec}  \approx \frac{n_e}{\sqrt{T_e^*}}      2.6 \cdot 10^{-19} \,\,\,\, {\rm s^{-1}}
,\end{equation}
\begin{equation}  \label{eq:gamma_ion}
\Gamma^{\rm ion}  \approx n_e A \frac{1}{X + \phi_{\rm ion}/{T_e^*}}\left(\frac{\phi_{\rm ion}}{T_e^*}\right)^K  e^{-\phi_{\rm ion}/T_e^*} \,\,\,\,  {\rm s^{-1}}
,\end{equation}
where 
$\phi_{\rm ion} = 13.6 eV$, 
$T_e^*$ is electron temperature in eV, and constants have values
$A = 2.91 \cdot 10^{-14}$, 
$K$ = 0.39, and $X$ = 0.232.

Eqs.~(\ref{eq:alpha_el}), (\ref{eq:gamma_rec}) and (\ref{eq:gamma_ion})
show the calculation
of $\alpha$, $\Gamma^{\rm ion}$ and $\Gamma^{\rm rec}$ in the SI unit system.
In  {\tt MPI-AMRVAC}, which uses a convenient non-dimensionalization, the most straightforward implementation was to 
first transform non-dimensional number densities and temperatures into the physical unit system defined for the simulation (which can be ``cgs'' or ``SI'' in {\tt MPI-AMRVAC})
by multiplying by their units $\bar{n}$ and $\bar{T}$ and further transform to the SI unit system, if needed.
After the calculation, the resulting quantities with physical SI units are again transformed
to the unit system of the simulation, then non-dimensionalized by dividing by the corresponding units:
$\bar{\alpha} = 1/(\bar{\rho} \bar{t})\,, \bar{\Gamma^{\rm ion}} = \bar{\Gamma^{\rm rec}} = 1/\bar{t}$.
\end{appendix}
\theendnotes
\end{document}